\documentclass[seceq]{ptptex}

\def\Vec#1{\mbox{\boldmath$#1$}}
\usepackage{graphics}

\title{
Boltzmann Equation for Relativistic Neutral Scalar Field 
in Non-equilibrium Thermo Field Dynamics
}

\author{
Yuichi \textsc{Mizutani}$^1$ and Tomohiro \textsc{Inagaki}$^2$
}

\inst{
  $^1$Department of Physics, Hiroshima University, 
 Higashi-Hiroshima, Hiroshima 739-8526, Japan \\
 $^2$ Information Media Center, Hiroshima University,
 Higashi-Hiroshima, Hiroshima 739-8521, Japan
}

\abst{
A relativistic neutral scalar field is investigated on the basis 
of the Schwinger-Dyson equation in the non-equilibrium thermo field 
dynamics. A time evolution equation for a distribution function is 
obtained from a diagonalization condition for the Schwinger-Dyson 
equation. An explicit expression of the time evolution equation is 
calculated in the $\lambda\phi^4$ interaction model at the 2-loop 
level. The Boltzmann equation is derived for the relativistic 
scalar field. We set a simple initial condition and numerically 
solve the Boltzmann equation and show the time evolution of the 
distribution function and the relaxation time. 
}

\begin{document}

\maketitle

\section{Introduction}
The relativity and the quantum field theory provide a general basis 
for understanding a model of elementary particles at high energy. 
The theory should be extended to include a statistical thermodynamics 
for macroscopic phenomena of many particles systems. 
Some formalism for thermal quantum field theories is proposed to 
evaluate in and out of equilibrium systems. 
The real-time formalism is necessary to include time evolution of 
the system. Y.~Takahashi and H.~Umezawa proposed the thermo field 
dynamics (TFD) which is the real-time formalism based on the canonical 
quantization \cite{finitetfd1,finitetfd2,umezawa1}. After that 
TFD has been applied to various physical systems. 

In 1985 T.~Arimitsu and H.~Umezawa extended TFD to a 
non-equilibrium system with spatially homogeneous 
distributions \cite{netfd0,arimitsu1}. It has been found that 
the time development of a particle number density is derived 
by the self-consistent renormalization condition 
\cite{umezawa1,CQNETFD1,CQNETFD2,CQNETFD3}. An equation 
with the structure of the Boltzmann equation is derived 
in the non-equilibrium TFD (NETFD). 
An alternative procedure to derive a Boltzmann-like equation 
has been proposed by evaluating the expectation value of the 
number operator in terms of a renormalized field in 
Refs.~\cite{matsumoto3,matsumoto5}. 
It is also obtained by diagonalizing the Green's function at 
the equal time limit instead of the thermal self-energy 
diagonalization scheme in Ref.~\cite{umezawa2}. 
TFD for spatially inhomogeneous non-equilibrium system has 
been studied in Refs.\cite{inhomo1,inhomo2,henning1}. 
NETFD successfully describes macroscopic 
phenomena in a low energy non-relativistic system. 

It is also quite interesting to make investigations various 
phenomena associated with non-equilibrium dynamics at high energy. 
As far as we know, only a little work has been reported in the 
study of NETFD for a high energy relativistic system. 
P.~Henning introduced independent weighting functions for 
relativistic fields with positive and negative energy and 
described an extended Bogoliubov transformation for thermal 
doublets \cite{henning1}. NETFD was applied for a relativistic 
scalar field with a four-point self-interaction in 
Ref.\cite{matsumoto4}. A renormalization condition is imposed 
on the thermal self-energy. Thus a Boltzmann-like equation is 
derived from the expression for the particle number density 
of the scalar field. 

In the present paper we would like to find a natural derivation of 
the self-consistency conditions available for relativistic scalar 
fields by using the Schwinger-Dyson (SD) equation. 
In Sec.~2 we briefly review NETFD for a scalar field and 
introduce the thermal Bogoliubov matrices whose parameter 
defines a particle number density. 
In Sec.~3 we evaluate the SD equation in NETFD. 
A Boltzmann equation is derived from a time 
evolution of a parameter in the thermal Bogoliubov matrices. 
In Sec.~4 a scalar field with a $\lambda \phi^4$ interaction 
is investigated according to the NETFD. 
We calculate the thermal self-energy at 2-loop level. 
In Sec.~5 we give an explicit expression for the Boltzmann 
equation in the $\lambda \phi^4$ interaction model. 
Solving the Boltzmann equation starting from a simple 
initial state numerically, we show the time evolution of 
the particle number density and the relaxation time. We 
discuss how the off-shell mode contributes the relaxation process. 
Some concluding remarks are given in Sec.~6. 

\section{Non-equilibrium thermo field dynamics for relativistic fields}
There are several formalism to introduce the thermal dynamics into 
the quantum field theory. In TFD the statistical average is replaced 
by an expectation value in a pure state called the thermal vacuum. 
The thermal vacuum is defined by extending the Fock space structure. 
A creation and an annihilation operators, $a^\dagger$, $a$ are doubled 
by introducing tilde operators, $\tilde{a}^\dagger$, $\tilde{a}$. 
Thus the commutation relations for the bosonic creation and 
annihilation operators are extended to be 
\begin{eqnarray}
&& [a_{\Vec{p}},a_{\Vec{q}}^\dagger]
=(2 \pi)^3 \delta^{(3)}(\Vec{p}-\Vec{q}),  \\
&& [\tilde{a}_{\Vec{p}}, \tilde{a}_{\Vec{q}}^\dagger ]
=(2 \pi)^3 \delta^{(3)}(\Vec{p}-\Vec{q}),  \\
&& {\rm others}=0.
\label{com0}
\end{eqnarray}
The time evolution of the non-tilde operators are generated by 
an ordinary non-thermal Hamiltonian, $H$, for the considering 
system. In a similar manner the time evolution for the tilde 
operators is described by the tilde conjugate Hamiltonian, 
$\tilde{H}$. In defining the tilde-Hamiltonian, $\tilde{H}$, 
we use the following tilde conjugation rules, 
\begin{eqnarray}
&& (A_1 A_2)^\sim=\tilde{A_1}\tilde{A_2},  \label{tilde1} \\
&& (c_1 A_1 + c_2 A_2)^\sim=c_1^\ast \tilde{A}_1 + c_2^\ast \tilde{A}_2,  \label{tilde2} \\
&& (\tilde{A})^\sim=A,	\label{tilde3}
\end{eqnarray}
where $c_1$ and $c_2$ are c-numbers, and $A_1$ and $A_2$ 
are arbitrary operators. The tilde-Hamiltonian, $\tilde{H}$, is 
constructed by only the tilde operators, $\tilde{a}$ and 
$\tilde{a}^\dagger$. 
The time evolution of both the non-tilde and tilde operators 
is described by the total Hamiltonian of the system, 
\begin{eqnarray}
\hat{H} \equiv H-\tilde{H}.	\label{hatH}
\end{eqnarray}

The Fock space is extended to the state space spanned by 
both the non-tilde and tilde creation operators. 
The thermal vacuum is defined by the thermal Bogoliubov 
transformation of the eigenstates of the Hamiltonian. 
Non-equilibrium degree of freedom can be also introduced 
in TFD through the thermal Bogoliubov transformation 
\cite{umezawa1,netfd0,Bogoliubov2}, 
\begin{eqnarray}
&& \xi_{p}^\alpha e^{-i\omega_p\cdot t}
 =B(n_{p}(t))^{\alpha \beta}a_{p}^\beta (t),	\nonumber \\ 
&& \bar{\xi}_{p}^\alpha e^{i \omega_p \cdot t}
 = \bar{a}_{p}^\beta(t) B^{-1}(n_{p}(t))^{\beta\alpha},
 \label{TBT0}
\end{eqnarray}
where $\omega_p = \sqrt[]{ \Vec{p}^2+m^2 }$ is the relativistic 
energy eigenvalue for bosonic fields with a momentum, $\Vec{p}$, and mass, $m$. 
The upper indices are defined by the thermal doublets notation, 
\begin{eqnarray}
&& a^{\alpha}_p=
\left(
	\begin{array}{c}
	a_p \\
	\tilde{a}_p^\dagger\\
	\end{array}
\right), 	\ \ \ \ 
 \bar{a}_p^\alpha=
\left(
	\begin{array}{cc}
	a_p^\dagger &
	-\tilde{a}_p \\
	\end{array}
\right), \\
&& \xi^{\alpha}_p=
\left(
	\begin{array}{c}
	\xi_p \\
	\tilde{\xi}_p^\dagger\\
	\end{array}
\right), 	\ \ \ \ 
 \bar{\xi}_p^\alpha=
\left(
	\begin{array}{cc}
	\xi_p^\dagger &
	-\tilde{\xi}_p \\
	\end{array}
\right).
\end{eqnarray}
It is assumed that the thermal Bogoliubov matrices, 
$B(n_{p}(t))^{\alpha \beta}$ and 
$B^{-1}(n_{p}(t))^{\alpha \beta}$, have the same 
forms with the ones in equilibrium, 
\begin{eqnarray}
&& B(n_{p}(t))^{\alpha \beta}=
\left(
	\begin{array}{cc}
	1+ n_{p}(t) & - n_{p}(t) \\
	-1     & 1 \\
	\end{array}
\right),			\nonumber \\
&& B^{-1}(n_{p}(t))^{\alpha \beta}=
\left(
	\begin{array}{cc}
	1 &  n_{p}(t) \\
	1 & 1+ n_{p}(t) \\
	\end{array}
\right).
\label{BMatrix}
\end{eqnarray}
The Bogoliubov parameter, $n_{p}(t)$, 
depends on time and the relativistic energy eigenvalue 
through the momentum, $\Vec{p}$. 
 
In NETFD the canonical quantization has not been 
fully established for relativistic fields yet 
although some attempts have been done\cite{henning1}. 
In the present paper it is assumed that the quantum correction 
is calculated by perturbative theory in the Fock space 
spanned by the $\xi$-oscillators. 
The thermal vacuum, $|\theta\rangle$, is defined by the transformed operator, $\xi_p$, 
\begin{eqnarray}
\langle \theta |\xi_p^\dagger=\xi_p |\theta\rangle=0,\ \ \ \ \ 
\langle \theta | \tilde{\xi}_p^\dagger= \tilde{\xi}_p |\theta \rangle=0.
\label{thermalVacuum}
\end{eqnarray}
The state, $|\theta\rangle$, is invariant under the tilde conjugation 
(\ref{tilde1}), (\ref{tilde2}) and (\ref{tilde3}), 
\begin{eqnarray}
&&(\langle \theta|)^{\sim}=\langle \theta|,\ \ \ \ \ 	
(| \theta \rangle)^{\sim}=|\theta \rangle. 
\end{eqnarray}
In equilibrium system the statistical thermal average is obtained by 
evaluating the expectation value under the thermal vacuum. 

A neutral scalar field is expanded by using the $a$-oscillators, i.e. 
inverse of the thermal Bogoliubov transformations (\ref{TBT0}) on 
the $\xi$-oscillators, 
\begin{eqnarray}
\phi(x) &=& \int \frac{d^3\Vec{p}}{(2 \pi)^3}\frac{1}{ \sqrt[]{2\omega_{p}}}
(a_{p}(t_x) {\rm e}^{ {\rm i} \Vec{p} \cdot \Vec{x}} 
+ a_{p}^\dagger (t_x)  {\rm e}^{-{\rm i} \Vec{p} \cdot \Vec{x} }), \\
\tilde{\phi}(x) &=& \int \frac{d^3\Vec{p}}{(2 \pi)^3}\frac{1}{\sqrt[]{ 2\omega_{p} }}
(\tilde{a}_{p}(t_x) {\rm e}^{-{\rm i} \Vec{p}\cdot \Vec{x}}
 + \tilde{a}_{p}^\dagger (t_x) {\rm e}^{ {\rm i}\Vec{p} \cdot \Vec{x} }).
\end{eqnarray}
From Eq.~(\ref{hatH}) the time dependence for the 
operators $a$ and $a^\dagger$ is identical to 
$\tilde{a}^\dagger$ and $\tilde{a}$, respectively. 
Thus the thermal doublets notation is defined for the neutral 
scalar fields by 
\begin{eqnarray}
\phi^\alpha (x) &=& \int \frac{d^3\Vec{p}}{(2 \pi)^3}\frac{1}{ \sqrt[]{2\omega_{p}}}
\left\{ 
\left(
  \begin{array}{c}
	a_{p}(t_x) \\
	\tilde{a}_{p}^\dagger (t_x)
  \end{array}
\right)
 {\rm e}^{ {\rm i} \Vec{p} \cdot \Vec{x}} 
+
\left(
  \begin{array}{c}
	a_{p}^\dagger (t_x) \\
	\tilde{a}_{p} (t_x)
  \end{array}
\right)
 {\rm e}^{-{\rm i} \Vec{p} \cdot \Vec{x} }
\right\} \nonumber \\
 &=&  \int \frac{d^3\Vec{p}}{(2 \pi)^3}\frac{1}{ \sqrt[]{2\omega_{p}}}
\left\{ a_{p}^\alpha (t_x) {\rm e}^{ {\rm i} \Vec{p} \cdot \Vec{x}} 
+(\tau_3 \bar{a}_{p}(t_x)^T)^{\alpha}  {\rm e}^{-{\rm i} \Vec{p} \cdot \Vec{x} }
\right\}, \\
\bar{\phi}^\alpha(x) &=& \int \frac{d^3\Vec{p}}{(2 \pi)^3}\frac{1}{\sqrt[]{ 2\omega_{p} }}
 \left\{
    \left(
	a_{p}^\dagger (t_x) \ \ 
	-\tilde{a}_{p} (t_x)
    \right)
 {\rm e}^{-{\rm i} \Vec{p}\cdot \Vec{x}}
 +
 \left(
	a_{p}(t_x) \ \ 
	-\tilde{a}_{p}^\dagger (t_x)
\right)
 {\rm e}^{ {\rm i}\Vec{p} \cdot \Vec{x} }
\right\} \nonumber \\
 &=& \int \frac{d^3\Vec{p}}{(2 \pi)^3}\frac{1}{\sqrt[]{ 2\omega_{p} }}
 \left\{	\bar{a}_{p}^\alpha (t_x) {\rm e}^{-{\rm i} \Vec{p}\cdot \Vec{x}}
 + (a_{p}(t_x)\tau_3)^\alpha {\rm e}^{ {\rm i}\Vec{p} \cdot \Vec{x}} \right\},
\end{eqnarray}
where $\tau_3$ is the third Pauli matrix acting on the thermal doublets. 

Below we express fields and propagators by the 
t-representation, which is defined by taking the spatial Fourier transform. 
A thermal propagator is defined by an expectation value of 
the time-ordered product of two fields in the thermal vacuum. 
It has a $2\times 2$ matrix form with respect to the thermal index. 
For a neutral scalar field it is defined by 
\begin{eqnarray}
 D^{\alpha\beta} (t_x-t_y;\Vec{p})
 \equiv \langle \theta| 
  T[\phi^\alpha (t_x;\Vec{p}) \bar{\phi}^\beta (t_y;\Vec{p}) ]
	|\theta \rangle ,
\end{eqnarray}
where $\alpha$ and $\beta$ are the thermal indices. 
Thus the thermal propagator is obtained by the thermal Bogoliubov 
transformation of the non-thermal one 
\cite{umezawa1,CQNETFD2,netfdH,thrmpro}, 
\begin{eqnarray}
 D^{\alpha\beta} (t_x-t_y;\Vec{p}) = B^{-1}(n_{p}(t_x))^{\alpha\gamma_1} 
D_{0,R}^{\gamma_1 \gamma_2} (t_x-t_y;\Vec{p}) 
B(n_{p}(t_y))^{\gamma_2\beta}  && \nonumber \\
  +\{\tau_3 B(n_{p}(t_x))^T \}^{\alpha\gamma_1} 
 D_{0,A}^{\gamma_1 \gamma_2} (t_x-t_y;\Vec{p})
 \{ B^{-1}(n_{p}(t_y))^T\tau_3 \}^{\gamma_2 \beta},&&	
\label{ThermPro0}
\end{eqnarray}
where $ D_{0,R}^{\gamma_1 \gamma_2} (t_x-t_y;\Vec{p})$ and 
$D_{0,A}^{\gamma_1 \gamma_2} (t_x-t_y;\Vec{p})$ are 
the retarded and the advanced parts for the non-thermal 
propagator which are given by $2\times 2$ matrices with thermal indices 
$\gamma_1$ and $\gamma_2$. 
The non-thermal propagators for the scalar field have diagonal forms. 
Each component in $ D_{0,R}$ and $D_{0,A}$ is given by 
\begin{eqnarray}
&& D_{0,R}^{11} (t_x-t_y;\Vec{p})=\theta(t_x-t_y) \frac{1}{2\omega_p}
{\rm e}^{-i \omega_p \cdot (t_x-t_y)}, \label{BoseThermPro1-1} \\
&& D_{0,R}^{22} (t_x-t_y;\Vec{p})=-\theta(t_y-t_x) \frac{1}{2\omega_p}
{\rm e}^{-i \omega_p\cdot (t_x-t_y)}, \label{BoseThermPro1-2} \\
&& D_{0,A}^{11} (t_x-t_y;\Vec{p})=\theta(t_y-t_x) \frac{1}{2\omega_p}
{\rm e}^{i \omega_p\cdot (t_x-t_y)}, \label{BoseThermPro1-3} \\
&& D_{0,A}^{22} (t_x-t_y;\Vec{p})=-\theta(t_x-t_y) \frac{1}{2\omega_p}
{\rm e}^{i \omega_p\cdot (t_x-t_y)}, 	
\label{BoseThermPro1}	\\
&&{\rm other \ components }=0. \nonumber
\end{eqnarray}

In NETFD we can choose the boundary condition for the thermal 
Bogoliubov matrices (\ref{BMatrix}) to derive the ordinary 
Feynman rules for relativistic scalar fields\cite{umezawa1,evans}. 
It is convenient for perturbative calculations. 
In TFD the thermal average of the particle number density is 
given by the expectation value of the number operator 
$a_{H,p}^\dagger a_{H,p}$ in the thermal vacuum, 
\begin{eqnarray}
n_p(t)=\langle \theta_H|a_{H,p}^\dagger(t) a_{H,p}(t) |\theta_H \rangle,
\end{eqnarray}
where  the fields and ground state with the lower index, $H$, are 
written in the Heisenberg representation. It coincides with the 
Bogoliubov parameter, $n_p$, in the Bogoliubov matrices (\ref{BMatrix}) 
in an equilibrium state at tree level. 
Since quantum corrections can induce instability of the thermal vacuum 
(\ref{thermalVacuum}), the Bogoliubov parameter does not always 
correspond to the particle number density out of equilibrium states. 
Below we assume that a stable vacuum is found by redefining the 
Bogoliubov parameter and the observed particle number density is 
obtained by the Bogoliubov parameter in the stable vacuum at the 
equal time limit \cite{umezawa1}. 

\section{Time evolution equation}
H.~Umezawa and Y.~Yamanaka have introduced the self-consistent 
renormalization condition to derive the Boltzmann equation in 
NETFD\cite{umezawa1,CQNETFD1,CQNETFD2,CQNETFD3}. 
The hat-Hamiltonian (\ref{hatH}) does not contain terms 
proportional to $\xi\tilde{\xi}$. However, such a term 
is induced through the quantum correction in the thermal 
self-energy out of equilibrium states. It breaks the 
condition (\ref{thermalVacuum}) for the thermal vacuum.
The terms proportional to $\xi\tilde{\xi}$ can be eliminated 
by the thermal Bogoliubov transformation. We impose that the 
terms proportional to $\xi\tilde{\xi}$ vanish at the equal 
time limit. The perturbed thermal Bogoliubov parameter can be 
fixed by this self-consistent renormalization condition. 
The Boltzmann equation appears as a consequence of 
this renormalization condition. 
It is also derived from diagonalizing 
the full propagator given by the Dyson equation 
at the equal time limit in non-relativistic quantum field theories 
\cite{umezawa2}. 
In this work we improve this diagonalization condition to 
be suitable for a relativistic scalar field. 

In TFD the SD equation for the scalar field is given by 
\begin{eqnarray}
&&D_H^{\alpha\beta} (t_x-t_y;\Vec{p})=
D_0^{\alpha\beta} (t_x-t_y;\Vec{p}) \nonumber \\ 
&&\ \ + \int dt_{z_1} dt_{z_2} \ D_0^{\alpha \gamma_1}(t_x-t_{z_1};\Vec{p})
i\Sigma^{\gamma_1 \gamma_2} (t_{z_1}-t_{z_2};\Vec{p})
D_H^{\gamma_2\beta}(t_{z_2}-t_{y};\Vec{p}),	\label{BoseSDeq0}
\end{eqnarray}
where $D_H^{\alpha\beta}(t_x-t_y;\Vec{p})$ 
denotes the full thermal propagator, 
$D_0^{\alpha\beta}(t_x-t_y;\Vec{p})$ the thermal 
propagator at tree level and 
$\Sigma^{\alpha \beta}(t_{z_1}-t_{z_2};\Vec{p})$ 
the full thermal self-energy. 
Another expression for this equation is 
\begin{eqnarray}
&&D_H^{\alpha\beta}(t_x-t_y;\Vec{p})=
D_0^{\alpha\beta}(t_x-t_y;\Vec{p}) \nonumber \\ 
&&\ \ + \int dt_{z_1} dt_{z_2} \ D_H^{\alpha \gamma_1}(t_x-t_{z_1};\Vec{p})
i\Sigma^{\gamma_1 \gamma_2}(t_{z_1}-t_{z_2};\Vec{p})
D_0^{\gamma_2\beta}(t_{z_2}-t_{y};\Vec{p}).	\label{BoseSDeq0-2}
\end{eqnarray}
We suppose that the full propagator can be decomposed in a similar 
form with the tree level one (\ref{ThermPro0}), 
\begin{eqnarray}
&&D_H^{\alpha\beta} (t_x-t_y;\Vec{p})=
B^{-1}(n_{H,p}(t_x))^{\alpha\gamma_1} 
D_{H,R}^{\gamma_1 \gamma_2} (t_x-t_y;\Vec{p}) 
B(n_{H,p}(t_y))^{\gamma_2\beta}  \nonumber \\
&&\ \ + \{ \tau_3 B(n_{H,p}(t_x))^T \}^{\alpha\gamma_1} 
 D_{H,A}^{\gamma_1 \gamma_2} (t_x-t_y;\Vec{p})
 \{ B^{-1}(n_{H,p}(t_y))^T\tau_3 \}^{\gamma_2 \beta},	\label{perturbedp}
\end{eqnarray}
where $n_{H,p}(t_x)$ and $n_{H,p}(t_y)$ are the Bogoliubov parameters 
acting on the full thermal propagator from the left- and the right-hand 
sides, respectively. For simplicity we omit the thermal indices and the 
momentum label $\Vec{p}$ in the thermal propagator and the Bogoliubov 
parameters below. 

We decompose the propagators in the SD equation 
(\ref{BoseSDeq0}) in accordance with Eqs.~(\ref{ThermPro0}) and 
(\ref{perturbedp}) to see the matrix structure of the thermal 
propagator. After the Klein-Gordon 
operator, $(\partial_{t_x}^2+\Vec{p}^2+m^2)$, 
is applied on the left to both sides, Eq.~(\ref{BoseSDeq0}) reads 
\begin{eqnarray}
&&(\partial_{t_x}^2+\Vec{p}^2+m^2)\Bigl[ B^{-1}(n_{H}(t_x)) 
D_{H,R} (t_x-t_y) 
B(n_{H}(t_y))  \nonumber \\
&&\ \ \ \ \ \ +  \tau_3 B(n_{H}(t_x))^T  
 D_{H,A} (t_x-t_y)
  B^{-1}(n_{H}(t_y))^T\tau_3 \Bigr]	\nonumber \\
&&=
(\partial_{t_x}^2 +\Vec{p}^2+m^2) \Bigl[ B^{-1}(n (t_x)) 
D_{0,R} (t_x-t_y)
B(n (t_y))  \nonumber \\
&&\ \ \ \ \ \ +  \tau_3 B(n (t_x))^T  
 D_{0,A} (t_x-t_y)
  B^{-1}(n (t_y))^T \tau_3  \Bigr]  \nonumber \\
 && \ \ \ + \int dt_{z_1} dt_{z_2} \ (\partial_{t_x}^2+\Vec{p}^2+m^2)
\Bigl[ B^{-1}(n (t_x)) 
D_{0,R} (t_x-t_{z_1})
B(n (t_{z_1}))  \nonumber \\
&&\ \ \ \ \ \ + \tau_3 B(n (t_x))^T 
 D_{0,A} (t_x-t_{z_1})
  B^{-1}(n (t_{z_1}))^T\tau_3  \Bigr]
i\Sigma( t_{z_1}-t_{z_2})	\nonumber \\
&&\ \ \ \times \Bigl[ B^{-1}(n_{H}(t_{z_2})) 
D_{H,R} (t_{z_2}-t_y) 
B(n_{H}(t_y))  \nonumber \\
&&\ \ \ \ \ \ \ \ \ + \tau_3 B(n_{H}(t_{z_2}))^T  
 D_{H,A} (t_{z_2}-t_y)
  B^{-1}(n_{H}(t_y))^T \tau_3  \Bigr],
\label{sd2}
\end{eqnarray}
where $n(t)$ is the Bogoliubov parameter for the non-perturbed 
operator included in the thermal propagator at tree level. 

The Bogoliubov matrices and the Klein-Gordon operator do not 
commute. Substituting the expression (\ref{BMatrix}) to the 
Bogoliubov matrices and inserting an identity operators, 
$B^{-1}(n_H)B(n_H)$, we obtain 
\begin{eqnarray}
&&B^{-1}(n_H(t_x))  (\partial_{t_x}^2+\Vec{p}^2+m^2) 
\left(
   \begin{array}{c}
	D_{H,R}^{11}(t_x-t_y)
    \ \ \ \  \ \ \  O_{R,prop1}(t_x,t_y) \\
	\ \ \ \ \ \ \ \   0 \ \ \  \ \ \ \ \ \ \ \ \ \ 
   D_{H,R}^{22}(t_x-t_y)  \\
   \end{array}
\right) B(n_{H}(t_y))	\nonumber \\
&&+ \tau_3 B(n_H(t_x))^T  (\partial_{t_x}^2+\Vec{p}^2+m^2) 
\left(
   \begin{array}{c}
	D_{H,A}^{11}(t_x-t_y)
	\ \ \ \ \ \ \ \ \ \ \  0 \ \ \ \ \    \\
     O_{A,prop1}(t_x,t_y) \ \ 
   D_{H,A}^{22}(t_x-t_y)  \\
   \end{array}
\right) B(n_{H}(t_y))^T\tau_3	\nonumber \\
&&-  \int dt_s \Biggl[
B^{-1}(n_H(t_x)) \left(
   \begin{array}{c}
	\frac{1}{2}\Sigma_R (t_x-t_s)  D_{H,R}^{11}(t_s-t_y) 
  \ \ \  g_{x1}(t_x,t_y,t_s)  \\
  ~~~~~~~~ 0  ~~~~~~~~~
  \frac{1}{2}\Sigma_A (t_x-t_s) D_{H,R}^{22}(t_s-t_y) \\
   \end{array}
\right) B(n_H(t_y)) \nonumber \\
&& + B^{-1}(n_H(t_x)) \left(
   \begin{array}{c}
	g_{x2}(t_x,t_y,t_s) \ \ \ \ 
  -\frac{1}{2}\Sigma_R(t_x-t_s) D_{H,A}^{22}(t_s-t_y)  \\   
 \!\!\! -\frac{1}{2}\Sigma_A (t_x-t_s) D_{H,A}^{11}(t_s-t_y)
  \ \ \ \ \ \ \ \ \ \ \ 0 \ \ \ \ \ \   \\
   \end{array}
\right) B^{-1}(n_H(t_y))^T \tau_3 \nonumber \\
&& + \tau_3 B(n_H(t_x))^T \left(
   \begin{array}{c}
  \ \ \ \ \ 0  \ \ \ \ \ \ \ \ \ 
  -\frac{1}{2} \Sigma_A (t_x-t_s) D_{H,R}^{22}(t_s-t_y) \\
	- \frac{1}{2} \Sigma_R (t_x-t_s)  D_{H,R}^{11}(t_s-t_y) 
  \ \ \  g_{x3}(t_x,t_y,t_s)  \\
   \end{array}
\right) B(n_H(t_y)) \nonumber \\
&& + \tau_3 B(n_H(t_x))^T \left(
   \begin{array}{c}
 \!\!\! \frac{1}{2} \Sigma_A (t_x-t_s) D_{H,A}^{11}(t_s-t_y)
  \ \ \ \ \ \ \ \ \ \ \ 0 \ \ \ \ \ \  \ \ \   \\
	g_{x4}(t_x,t_y,t_s) \ \ \ \ 
  \frac{1}{2} \Sigma_R(t_x-t_s) D_{H,A}^{22}(t_s-t_y)  \\
   \end{array}
\right) B(n_H(t_y))^T \tau_3 \Biggr] \nonumber \\
&& =-i \delta (t_x-t_y),		\label{BoseSDeq2x}
\end{eqnarray}
where the thermal self-energies for the retarded and the advanced 
propagators, $\Sigma_R$ and $\Sigma_A$, are defined by 
\begin{eqnarray}
\Sigma_R \equiv \Sigma^{11}+\Sigma^{12} = \Sigma^{21}+\Sigma^{22},
 \ \ \ \ 
\Sigma_A \equiv  \Sigma^{11}-\Sigma^{21} = \Sigma^{22}-\Sigma^{12}.
\label{SelfERA}
\end{eqnarray}
The off-diagonal elements, $O_{R(A),prop1}$, in the propagator (\ref{BoseSDeq2x}) 
are defined to satisfy the following equations,  
\begin{eqnarray}
&&(\partial_{t_x}^2+\Vec{p}^2+m^2)O_{R,prop1}(t_x,t_y)	\nonumber \\
&&\ \ \ \equiv \ddot{n}_H(t_x) D_{H,R}^{22}(t_x-t_y)
	+ 2 \dot{n}_H(t_x)(\partial_{t_x}D_{H,R}^{22}(t_x-t_y))	\nonumber \\
&&\ \ \ \ \ \ - \ddot{n} (t_x)D_{0,R}^{22}(t_x-t_y) 
	- 2 \dot{n}(t_x) (\partial_{t_x}D_{0,R}^{22}(t_x-t_y)),	\label{BSENDx1-1}
	\\
&&(\partial_{t_x}^2+\Vec{p}^2+m^2)O_{A,prop1}(t_x,t_y)	\nonumber \\
&&\ \ \ \equiv  - \ddot{n}_{H}(t_x) D_{H,A}^{11}(t_x-t_y)
	-2 \dot{n}_H(t_x) (\partial_{t_x}D_{H,A}^{11}(t_x-t_y))	\nonumber \\
&&\ \ \ \ \ \ + \ddot{n}(t_x) D_{0,A}^{11}(t_x-t_y)
	+2 \dot{n}(t_x) (\partial_{t_x}D_{0,A}^{11}(t_x-t_y)),	\label{BSENDx1-2}
\end{eqnarray}
We define the off-diagonal elements, $g_{x1} \sim g_{x4}$, by 
\begin{eqnarray}
&&g_{x1}(t_x,t_y,t_s)	\nonumber \\
&&\equiv\frac{1}{2} \Bigl\{ \Sigma^{12}(t_x-t_s)
 + h_{-} (t_x,t_s) \Bigr\} D_{H,R}^{22}(t_s-t_y)	\nonumber \\
&& + \int_{-\infty}^{\infty} dt_{z}\Biggl[ \Big\{ 
	 \ddot{n}(t_x) D_{0,R}^{22}(t_x-t_s) 
	+2 \dot{n}(t_x) (\partial_{t_x}D_{0,R}^{22}(t_x-t_s)) \Big\} \nonumber \\
&& \ \ \ \times \ i \Sigma_{A} (t_s-t_z) 
  D_{H,R}^{22}(t_s-t_y)  \Biggl],	\label{BSENDx2-1} \\
&&g_{x2}(t_x,t_y,t_s)	\nonumber \\
&&\equiv  \frac{1}{2} \Bigl\{ \Sigma^{11}(t_x-t_s)
+ h_{+} (t_x,t_s) \Bigr\} D_{H,A}^{11}(t_s-t_y)	\nonumber \\
&& - \int_{-\infty}^{\infty} dt_{z}\Biggl[ \Big\{ 
	 \ddot{n}(t_x) D_{0,R}^{22}(t_x-t_s) 
	+2 \dot{n}(t_x) (\partial_{t_x}D_{0,R}^{22}(t_x-t_s))  \Big\} \nonumber \\
&& \ \ \ \times \ i \Sigma_{A} (t_s-t_z)
  D_{H,A}^{11}(t_z-t_y) \Biggl],	\label{BSENDx2-2} \\
&& g_{x3}(t_x,t_y,t_s)	\nonumber \\
&& \equiv -\frac{1}{2} \Bigl\{ \Sigma^{22}(t_x-t_s)
+ h_{+} (t_x,t_s) \Bigr\} D_{H,R}^{22}(t_s-t_y)	\nonumber \\
&& + \int_{-\infty}^{\infty} dt_{z}\Biggl[ \Big\{ 
	 \ddot{n}(t_x) D_{0,A}^{11}(t_x-t_s) 
	+2 \dot{n}(t_x) (\partial_{t_x}D_{0,A}^{11}(t_x-t_s)) \Big\} \nonumber \\
&& \ \ \ \times \ i \Sigma_{A} (t_s-t_z) 
  D_{H,R}^{22}(t_s-t_y)  \Biggl],	\label{BSENDx2-3} \\
&&g_{x4}(t_x,t_y,t_s)	\nonumber \\
&&\equiv -\frac{1}{2} \Bigl\{ \Sigma^{21}(t_x-t_s)
+ h_{-} (t_x,t_s) \Bigr\} D_{H,A}^{11}(t_s-t_y)	\nonumber \\
&& - \int_{-\infty}^{\infty} dt_{z}\Biggl[ \Big\{ 
	 \ddot{n}(t_x) D_{0,A}^{11}(t_x-t_s) 
	+2 \dot{n}(t_x) (\partial_{t_x}D_{0,A}^{11}(t_x-t_s)) \Big\} \nonumber \\
&& \ \ \ \times \ i \Sigma_{A} (t_s-t_z) 
D_{H,A}^{11}(t_s-t_y)  \Biggl],	\label{BSENDx2-4}
\end{eqnarray}
with
\begin{eqnarray}
&& h_{-} (t,t^\prime) \equiv n_H(t^\prime) \Sigma_{R}(t - t^\prime)
  - n_H(t) \Sigma_{A}(t - t^\prime ), \\ 
&& h_{+} (t,t^\prime) \equiv n_H(t^\prime) \Sigma_{R}(t - t^\prime)
  + n_H(t) \Sigma_{A}(t - t^\prime).
\end{eqnarray}
It should be noticed that the first and the second derivatives of 
the Bogoliubov parameters appear in Eq.~(\ref{BoseSDeq2x}) through the 
off-diagonal elements, $O_{R(A),prop1}$, and $g_{x1} \sim g_{x4}$. 

We also rewrite the SD equation (\ref{BoseSDeq0-2}). 
Applying the Klein-Gordon operator from right to both sides of 
Eq.~(\ref{BoseSDeq0-2}), we obtain 
\begin{eqnarray}
&&\Bigl[ B^{-1}(n_{H}(t_x)) 
D_{H,R} (t_x-t_y) 
B(n_{R}(t_y))  \nonumber \\
&&\ \ +  \tau_3 B(n_{H}(t_x))^T  
 D_{H,A} (t_x-t_y)
  B^{-1}(n_{H}(t_y))^T\tau_3  \Bigr]
(\overleftarrow{\partial}_{t_y}^2+\Vec{p}^2+m^2)	\nonumber \\
&&=
 \Bigl[ B^{-1}(n (t_x)) 
D_{0,R} (t_x-t_y)
B(n (t_y))  \nonumber \\
&&\ \ +  \tau_3 B(n (t_x))^T  
 D_{0,A} (t_x-t_y)
  B^{-1}(n (t_y))^T \tau_3  \Bigr]
(\overleftarrow{\partial}_{t_y}^2+\Vec{p}^2+m^2)  \nonumber \\
 && \ + \int dt_{z_1} dt_{z_2} 
\Bigl[ B^{-1}(n_H (t_x)) 
D_{H,R} (t_x-t_{z_1})
B(n_H (t_{z_1}))  \nonumber \\
&&\ \ \ \ +  \tau_3 B(n_H (t_x))^T  
 D_{H,A} (t_x-t_{z_1})
  B^{-1}(n_H (t_{z_1}))^T\tau_3  \Bigr]	\nonumber \\
&&\ \ \times i\Sigma( t_{z_1}-t_{z_2}) \Bigl[ B^{-1}(n(t_{z_2})) 
D_{0,R} (t_{z_2}-t_y) 
B(n (t_y))  \nonumber \\
&&\ \ \ \ +  \tau_3 B(n(t_{z_2}))^T  
 D_{0,A} (t_{z_2}-t_y)
  B^{-1}(n (t_y))^T \tau_3  \Bigr]
(\overleftarrow{\partial}_{t_y}^2+\Vec{p}^2+m^2).
\label{sd4}	
\end{eqnarray}
Substituting Eq.~(\ref{BMatrix}) to this equation, Eq.~(\ref{sd4}) reads 
\begin{eqnarray}
&&B^{-1}(n_H(t_x))  
\left(
   \begin{array}{c}
	D_{H,R}^{11}(t_x-t_y)
    \ \ \ \  \ \ \  O_{R,prop2}(t_x,t_y) \\
	\ \ \ \ \ \ \ \   0 \ \ \  \ \ \ \ \ \ \ \ \ \ 
   D_{H,R}^{22}(t_x-t_y)  \\
   \end{array}
\right) (\overleftarrow{\partial}_{t_y}^2+\Vec{p}^2+m^2)
 B(n_{H}(t_y))	\nonumber \\
&&+ \tau_3 B(n_H(t_x))^T 
\left(
   \begin{array}{c}
	D_{H,A}^{11}(t_x-t_y)
	\ \ \ \ \ \ \ \ \ \ \  0 \ \ \ \ \    \\
     O_{A,prop2}(t_x,t_y) \ \ 
   D_{H,A}^{22}(t_x-t_y)  \\
   \end{array}
\right)
  (\overleftarrow{\partial}_{t_y}^2+\Vec{p}^2+m^2) B^{-1}(n_{H}(t_y))^T\tau_3
	\nonumber \\
&& -  \int dt_s \Biggl[
B^{-1}(n_H(t_x)) \left(
   \begin{array}{c}
	\frac{1}{2} D_{H,R}^{11}(t_x-t_s) \Sigma_R (t_s-t_y) 
  \ \ \  g_{y1}(t_x,t_y,t_s)  \\
  ~~~~~~~~~ 0  ~~~~~~~~
  \frac{1}{2} D_{H,R}^{22}(t_x-t_s) \Sigma_A (t_s-t_y) \\
   \end{array}
\right) B(n_H(t_y)) \nonumber \\
&& + B^{-1}(n_H(t_x)) \left(
   \begin{array}{c}
	g_{y2}(t_x,t_y,t_s) \ \ \ \ 
  -\frac{1}{2} D_{H,R}^{11}(t_x-t_s) \Sigma_R(t_s-t_y)   \\   
 \!\!\! - \frac{1}{2} D_{H,R}^{22}(t_x-t_s) \Sigma_A (t_s-t_y)
  ~~~~~~~~~ 0 ~~~~~~~   \\
   \end{array}
\right) B^{-1}(n_H(t_y))^T \tau_3 \nonumber \\
&& + \tau_3 B(n_H(t_x))^T \left(
   \begin{array}{c}
  ~~~~~~~ 0  ~~~~~~~
  - \frac{1}{2} D_{H,A}^{11}(t_x-t_s) \Sigma_A (t_s-t_y) \\
	- \frac{1}{2} D_{H,A}^{22}(t_x-t_s) \Sigma_R (t_s-t_y) 
  \ \ \  g_{y3}(t_x,t_y,t_s)  \\
   \end{array}
\right) B(n_H(t_y)) \nonumber \\
&& + \tau_3 B(n_H(t_x))^T \left(
   \begin{array}{c}
 \!\!\! \frac{1}{2} D_{H,A}^{11}(t_x-t_s) \Sigma_A (t_s-t_y) 
  ~~~~~~~~~~ 0 ~~~~~~~   \\
	g_{y4}(t_x,t_y,t_s) \ \ \ \ 
  \frac{1}{2} D_{H,A}^{22}(t_x-t_s) \Sigma_R (t_s-t_y)  \\
   \end{array}
\right) B^{-1}(n_H(t_y))^T \tau_3 \Biggr] \nonumber \\
&& =-i \delta (t_x-t_y),		\label{BoseSDeq2y}
\end{eqnarray}
where the off-diagonal elements, $O_{R(A),prop2}$, are defined to satisfy 
\begin{eqnarray}
&&O_{R,prop2}(t_x,t_y)(\overleftarrow{\partial}_{t_y}^2+\Vec{p}^2+m^2) 
	\nonumber \\
&&\ \ \ = -\ddot{n}_H(t_y) D_{H,R}^{11}(t_x-t_y)
	- 2 \dot{n}_H(t_y)(\partial_{t_y}D_{H,R}^{11}(t_x-t_y))	\nonumber \\
&&\ \ \ \ \ \ + \ddot{n} (t_y)D_{0,R}^{11}(t_x-t_y) 
	+ 2 \dot{n}(t_y) (\partial_{t_y}D_{0,R}^{11}(t_x-t_y)),	\label{BSENDy-1}
	\\
&&O_{A,prop2}(t_x,t_y)(\overleftarrow{\partial}_{t_y}^2+\Vec{p}^2+m^2) 
	\nonumber \\
&&\ \ \ =   \ddot{n}_H(t_y) D_{H,A}^{22}(t_x-t_y)
	+ 2 \dot{n}_H(t_y) (\partial_{t_y}D_{H,A}^{22}(t_x-t_y))	\nonumber \\
&&\ \ \ \ \ \ - \ddot{n}(t_x) D_{0,A}^{22}(t_x-t_y)
	- 2 \dot{n}(t_x) (\partial_{t_y}D_{0,A}^{22}(t_x-t_y)),	\label{BSENDy-2}
\end{eqnarray}
and the off-diagonal elements, $g_{y1} \sim g_{y4}$, are 
\begin{eqnarray}
&&g_{y1}(t_x,t_y,t_s)	\nonumber \\
&&=\frac{1}{2} D_{H,R}^{11}(t_x-t_s)
  \Bigl\{ \Sigma^{12}(t_s-t_y)
+ h_{-} (t_s,t_y) \Bigr\}	\nonumber \\
&& - \int_{-\infty}^{\infty} dt_{z}\Biggl[ 
 D_{H,R}^{11}(t_x-t_z) i \Sigma_{R} (t_z-t_s) \nonumber \\
&& \ \ \ \times \Big\{ 
	 \ddot{n}(t_y) D_{0,R}^{11}(t_s-t_y) 
	+2 \dot{n}(t_y) (\partial_{t_y}D_{0,R}^{11}(t_s-t_y)) \Big\}
  \Biggl],	\label{BSENDy2-1} \\
&&g_{y2}(t_x,t_y,t_s)	\nonumber \\
&&=  \frac{1}{2} D_{H,R}^{11}(t_x-t_s)
\Bigl\{ \Sigma^{11}(t_s-t_y)
+ h_{+} (t_s,t_y) \Bigr\}	\nonumber \\
&& - \int_{-\infty}^{\infty} dt_{z}\Biggl[ 
 D_{H,R}^{11}(t_x-t_z) i \Sigma_{R} (t_z-t_s) \nonumber \\
&& \ \ \ \times \Big\{
	\ddot{n}(t_y) D_{0,A}^{22}(t_s-t_y) 
	+2 \dot{n}(t_y) (\partial_{t_y}D_{0,A}^{22}(t_s-t_y))  \Big\}
    \Biggl],	\label{BSENDy2-2} \\
&& g_{y3}(t_x,t_y,t_s)	\nonumber \\
&& = -\frac{1}{2} D_{H,A}^{22} (t_x-t_s)
  \Bigl\{ \Sigma^{22}(t_s-t_y)
+ h_{+} (t_s,t_y) \Bigr\}	\nonumber \\
&& + \int_{-\infty}^{\infty} dt_{z}	\Biggl[
	 D_{H,A}^{22}(t_x-t_z) i \Sigma_{R} (t_z-t_s) \nonumber \\
&& \ \ \ \times \Big\{
	\ddot{n}(t_y) D_{0,R}^{11}(t_s-t_y) 
	+2 \dot{n}(t_y) (\partial_{t_y}D_{0,R}^{11}(t_s-t_y)) \Big\}
  \Biggl],	\label{BSENDy2-3} \\
&&g_{y4}(t_x,t_y,t_s)	\nonumber \\
&&= -\frac{1}{2} D_{H,A}^{22} (t_x-t_s)
\Bigl\{ \Sigma^{21}(t_s-t_y)
+ h_{-} (t_s,t_y) \Bigr\}	\nonumber \\
&& + \int_{-\infty}^{\infty} dt_{z}\Biggl[ 
 D_{H,A}^{22}(t_x-t_z) i \Sigma_{R} (t_z-t_s) \nonumber \\
&& \ \ \ \times \Bigl\{
	\ddot{n}(t_y) D_{0,A}^{22}(t_s-t_y)
	+ 2 \dot{n}(t_y) (\partial_{t_y}D_{0,A}^{22}(t_s-t_y)) \Bigr\}
	\Biggr].	\label{BSENDy2-4}
\end{eqnarray}  

For non-relativistic fields the particle number conservation law 
is derived from the difference between the SD equation 
applying the differential operator, $i\partial_{t_x}+\Vec{p}^2/2m$, 
on the left and right sides at the equal 
time limit \cite{Kadanoff}. We adapt the procedure to relativistic 
scalar fields. Subtracting the Eq.~(\ref{BoseSDeq2y}) from 
Eq.~(\ref{BoseSDeq2x}), we obtain 
\begin{eqnarray}
&&B^{-1}(n_H(t_x))\Biggl[  (\partial_{t_x}^2+\Vec{p}^2+m^2) 
\left(
   \begin{array}{c}
	D_{H,R}^{11}(t_x-t_y)
    \ \ \ \  \ \ \  O_{R,prop1}(t_x,t_y) \\
	\ \ \ \ \ \ \ \   0 \ \ \  \ \ \ \ \ \ \ \ \ \ 
   D_{H,R}^{22}(t_x-t_y)  \\
   \end{array}
\right) \nonumber \\
&&\ \ \ - \left(
   \begin{array}{c}
	D_{H,R}^{11}(t_x-t_y)
    \ \ \ \  \ \ \  O_{R,prop2}(t_x,t_y) \\
	\ \ \ \ \ \ \ \   0 \ \ \  \ \ \ \ \ \ \ \ \ \ 
   D_{H,R}^{22}(t_x-t_y)  \\
   \end{array}
\right) (\overleftarrow{\partial}_{t_y}^2+\Vec{p}^2+m^2) \nonumber \\
&& \ \ -  \int dt_s \Biggl\{
 \left(
   \begin{array}{c}
	\frac{1}{2}\Sigma_R (t_x-t_s)  D_{H,R}^{11}(t_s-t_y) 
  \ \ \  g_{x1}(t_x,t_y,t_s)  \\
  ~~~~~~~~ 0  ~~~~~~~~~
  \frac{1}{2} \Sigma_A (t_x-t_s) D_{H,R}^{22}(t_s-t_y) \\
   \end{array}
\right)  \nonumber \\
&&\ \ \ \ \  - \left(
   \begin{array}{c}
	\frac{1}{2} D_{H,R}^{11}(t_x-t_s) \Sigma_R (t_s-t_y) 
  \ \ \  g_{y1}(t_x,t_y,t_s)  \\
  ~~~~~~~~ 0  ~~~~~~~~~
  \frac{1}{2} D_{H,R}^{22}(t_x-t_s) \Sigma_A (t_s-t_y) \\
   \end{array}
\right) \Biggr\}
\Biggr] B(n_{H}(t_y))	\nonumber \\
&&+ \tau_3 B(n_H(t_x))^T \Biggl[ (\partial_{t_x}^2+\Vec{p}^2+m^2) 
\left(
   \begin{array}{c}
	D_{H,A}^{11}(t_x-t_y)
	~~\ \ \ \ \ \  \ \ \  0 \ \ \ \ \ ~~~  \\
     O_{A,prop1}(t_x,t_y) \ \ 
   D_{H,A}^{22}(t_x-t_y)  \\
   \end{array}
\right) \nonumber \\
&&\ \ \  - \left(
   \begin{array}{c}
	D_{H,A}^{11}(t_x-t_y)
	\ \ \ \ \ \ \ \ \ \ \  0 \ \ \ \ \    \\
     O_{A,prop2}(t_x,t_y) \ \ 
   D_{H,A}^{22}(t_x-t_y)  \\
   \end{array}
\right)
  (\overleftarrow{\partial}_{t_y}^2+\Vec{p}^2+m^2) \nonumber \\
&&\ \  - \int dt_s \Biggl\{
 \left(
   \begin{array}{c}
 \!\!\! \frac{1}{2} \Sigma_A (t_x-t_s) D_{H,A}^{11}(t_s-t_y)
  \ \ \ \ \ \ \ \ \ \ \ 0 \ \ \ \ \ \  \ \ \   \\
	g_{x4}(t_x,t_y,t_s) \ \ \ \ 
  \frac{1}{2} \Sigma_R(t_x-t_s) D_{H,A}^{22}(t_s-t_y)  \\
   \end{array}
\right) \nonumber \\
&&\ \ \ \ -
 \left(
   \begin{array}{c}
 \!\!\! \frac{1}{2} D_{H,A}^{11}(t_x-t_s) \Sigma_A (t_s-t_y) 
  \ \ \ \ \ \ \ \ \ \ \ 0 \ \ \ \ \ \  \ \ \   \\
	g_{y4}(t_x,t_y,t_s) \ \ \ \ 
  \frac{1}{2} D_{H,A}^{22}(t_x-t_s) \Sigma_R (t_s-t_y)  \\
   \end{array}
\right)
 \Biggr\}  \Biggr] B^{-1}(n_{H}(t_y))^T\tau_3	\nonumber \\
&& - B^{-1}(n_H(t_x)) \int dt_s \Biggl[ \left(
   \begin{array}{c}
	g_{x2}(t_x,t_y,t_s) \ \ \ \ 
  - \frac{1}{2} \Sigma_R(t_x-t_s) D_{H,A}^{22}(t_s-t_y)  \\   
 \!\!\! - \frac{1}{2} \Sigma_A (t_x-t_s) D_{H,A}^{11}(t_s-t_y)
  \ \ \ \ \ \ \ \ \ \ \ 0 \ \ \ \ \ \  \ \ \   \\
   \end{array}
\right) \nonumber \\
&&\ \ \ - \left(
   \begin{array}{c}
	g_{y2}(t_x,t_y,t_s) \ \ \ \ 
  - \frac{1}{2} D_{H,R}^{11}(t_x-t_s) \Sigma_R(t_s-t_y)   \\   
 \!\!\! - \frac{1}{2} D_{H,R}^{22}(t_x-t_s) \Sigma_A (t_s-t_y)
  \ \ \ \ \ \ \ \ \ \ \ 0 \ \ \ \ \ \  \ \ \   \\
   \end{array}
\right)
\Biggr] B^{-1}(n_H(t_y))^T \tau_3 \nonumber \\
&& - \tau_3 B(n_H(t_x))^T \int dt_s \Biggl[ \left(
   \begin{array}{c}
  ~~~~~~~~ 0  ~~~~~~~
  - \frac{1}{2} \Sigma_A (t_x-t_s) D_{H,R}^{22}(t_s-t_y) \\
	- \frac{1}{2} \Sigma_R (t_x-t_s)  D_{H,R}^{11}(t_s-t_y) 
  \ \ \  g_{x3}(t_x,t_y,t_s)  \\
   \end{array}
\right) \nonumber \\
&&\ \ \ - \left(
   \begin{array}{c}
  ~~~~~~~~ 0 ~~~~~~~~
  - \frac{1}{2} D_{H,A}^{11}(t_x-t_s) \Sigma_A (t_s-t_y) \\
	- \frac{1}{2} D_{H,A}^{22}(t_x-t_s) \Sigma_R (t_s-t_y) 
  \ \ \  g_{y3}(t_x,t_y,t_s)  \\
   \end{array}
\right) \Biggr] B(n_H(t_y))  
 = 0.		\label{BoseSDeqx-y}
\end{eqnarray}
Each elements of the matrices between the Bogoliubov matrices 
$\tau_3B^T(\cdots)B$ and $B^{-1}(\cdots)B^{-1T}\tau_3$ 
and the diagonal element in 
$B^{-1}(\cdots)B$ and $\tau_3B^T(\cdots)B^{-1T}\tau_3$ 
satisfies trivial equation, $0=0$, at the equal time limit, $t_x \rightarrow t_y$, 
in the case of the relativistic 
scalar field with a four-point self-interaction, as is shown in the next section. 
The remaining off-diagonal elements should satisfy 
\begin{eqnarray}
&& \lim_{t_x\rightarrow t_y}\Bigl[ (\partial_{t_x}^2+\Vec{p}^2+m^2)O_{R,prop1}(t_x,t_y) 
 - O_{R,prop2}(t_x,t_y) (\overleftarrow{\partial}_{t_y}^2+\Vec{p}^2+m^2) \nonumber \\
&&-\int dt_s \ \Bigl\{ g_{x1}(t_x,t_y,t_s) 
 - g_{y1}(t_x,t_y,t_s) \Bigr\} \Bigr]
=0, 	\label{BoseSDeq3r}  \\
&& \lim_{t_x\rightarrow t_y} \Bigl[ (\partial_{t_x}^2+\Vec{p}^2+m^2)O_{A,prop1}(t_x,t_y) 
 - O_{A,prop2}(t_x,t_y) (\overleftarrow{\partial}_{t_y}^2+\Vec{p}^2+m^2) \nonumber \\
&&- \int dt_s \ \Bigl\{ g_{x4}(t_x,t_y,t_s) 
 - g_{y4}(t_x,t_y,t_s) \Bigr\} \Bigr] 
=0. 	\label{BoseSDeq3a}
\end{eqnarray}
Substituting Eqs.~(\ref{BSENDx1-1}), (\ref{BSENDx2-1}), (\ref{BSENDy-1}) 
and (\ref{BSENDy2-1}) into Eq.~(\ref{BoseSDeq3r}), we obtain 
\begin{eqnarray}
&& \lim_{t_x\rightarrow t_y} \Bigl( \ddot{n}_H (t_x) D_{H,R}^{22}(t_x-t_y) 
- \ddot{n}(t_x) D_{0,R}^{22}(t_x-t_y)\nonumber \\
&&\  + \ddot{n}_H(t_y) D_{H,R}^{11}(t_x-t_y) 
 - \ddot{n}(t_y) D_{0,R}^{11}(t_x-t_y) \nonumber \\
&&\ +2\dot{n}_{H}(t_x) (\partial_{t_x}D_{H,R}^{22}(t_x-t_y))
	- 2\dot{n}(t_x)(\partial_{t_x}D_{0,R}^{22}(t_x-t_y))
	\nonumber \\
&&\ \ + 2\dot{n}_{H}(t_y) (\partial_{t_y}D_{H,R}^{11}(t_x-t_y))
	-2 \dot{n}(t_y) (\partial_{t_y}D_{0,R}^{11}(t_x-t_y))	\nonumber \\
&& + \frac{i}{2}\int_{-\infty}^{\infty} dt_s
  \Big[ \Bigl\{i\Sigma^{12}(t_x-t_s)
+ i h_{-} (t_x,t_s) \Bigr\} D_{H,R}^{22}(t_s-t_y)	\nonumber \\
&&\ \ \ -D_{H,R}^{11}(t_x-t_s)
  \Bigl\{ i\Sigma^{12}(t_s-t_y)
+ i h_{-} (t_s,t_y) \Bigr\} \Bigr] \Bigr)
=0.  \label{BoseSDeq4a}
\end{eqnarray}
This equation describes the time evolution of the Bogoliubov parameter, $n(t)$. 

We write the differences between the perturbed and the unperturbed Bogoliubov 
parameter as 
\begin{eqnarray}
\nu (t) \equiv n_{H}(t) - n(t).		\label{BoseBogoliP1}
\end{eqnarray}

Substituting Eqs.~(\ref{DHeqltim1})-(\ref{DHeqltim4}) to Eq.~(\ref{BoseSDeq4a}), 
a delta function , $\delta(t_x-t_y)$, appears from the first derivative of the propagator. 
It diverges at the equal time limit. 
The divergence is canceled out in Eq.~(\ref{BoseSDeq4a}). 
The ordinary Boltzmann equation does not contain terms 
proportional to the second derivative of the 
thermal Bogoliubov parameter, $\ddot{\nu}$. 
These terms in Eq.~(\ref{BoseSDeq4a}) are canceled out 
at the equal time limit. 
Therefore the equation (\ref{BoseSDeq4a}) is simplified to a Markovian equation, 
\begin{eqnarray}
\dot{\nu} (t_x)
 &=& -\frac{1}{2} \lim_{t_x\rightarrow t_y}
\int dt_s  \Big[ \Bigl\{ i \Sigma^{12}(t_x-t_s)
+ i h_{-} (t_x,t_s) \Bigr\} D_{H,R}^{22}(t_s-t_y)	\nonumber \\
&&\ \ \  - D_{H,R}^{11}(t_x-t_s)
  \Bigl\{ i \Sigma^{12}(t_s-t_y)
 + i h_{-} (t_s,t_y) \Bigr\} \Bigr].	  \label{BltzEqR-2}
\end{eqnarray}
In a similar manner Eq.~(\ref{BoseSDeq3a}) reduces to a 
Markovian equation for the advanced propagator, 
\begin{eqnarray}
\dot{\nu}(t_x)
&=& - \frac{1}{2} \lim_{t_x \rightarrow t_y} \int dt_s  \Bigl[
  \Bigl\{ i \Sigma^{21}(t_x-t_s)
+ i h_{-} (t_x,t_s) \Bigr\} D_{H,A}^{11}(t_s-t_y)	\nonumber \\
&&\ \ \  - D_{H,A}^{22} (t_x-t_s)
 \Bigl\{ i \Sigma^{21}(t_s-t_y)
+ i h_{-} (t_s,t_y) \Bigr\}
	\Bigr]. \label{BltzEqA-2}
\end{eqnarray}
Therefore the time evolution of the Bogoliubov parameter is determined by 
solving Eqs.~(\ref{BltzEqR-2}) and (\ref{BltzEqA-2}). 

\section{Boltzmann equation in a $\lambda\phi^4$ theory}
It is expected that the time evolution of the number distribution 
is given by the Boltzmann equation in NETFD. 
In this section we show that the Boltzmann equation is 
obtained from Eqs.~(\ref{BltzEqR-2}) and (\ref{BltzEqA-2}). 
We consider a neutral scalar field with a four-point 
self-interaction in (1+2) dimensions, for simplicity. To derive the 
Boltzmann equation we calculate the self-energy and evaluate 
the time evolution equations (\ref{BltzEqR-2}) and (\ref{BltzEqA-2}). 
We start from the Lagrangian density 
\begin{eqnarray}
{\cal L}(x)=\frac{1}{2}(\partial_\mu\phi(x))(\partial^\mu\phi(x)) 
-\frac{1}{2}m^2\phi(x)^2 - \frac{\lambda}{4!}\phi(x)^4. \nonumber \\
	\label{lagrangian}
\end{eqnarray}
In TFD the total Lagrangian density, $\hat{{\cal L}}$, for non-tilde and tilde fields is given by 
\begin{eqnarray}
\hat{{\cal L}}(x)\equiv {\cal L}(x)-\tilde{{\cal L}}(x),	\label{hatL}
\end{eqnarray}
where $\tilde{\cal L}$ is the tilde conjugate of the Lagrangian 
density ${\cal L}$. In this model the self-energy at 1-loop level 
can not contribute to the right-hand sides in Eqs.~(\ref{BltzEqR-2}) 
and (\ref{BltzEqA-2}). 
We have to calculate the self-energy at the 2-loop level.

\begin{figure}[b]
	\begin{center}
	\includegraphics{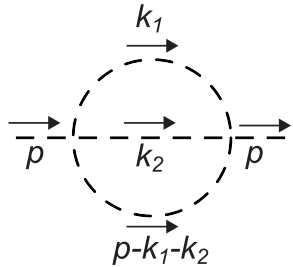}
	\end{center}
	\caption{2-loop thermal self-energy in $\lambda \phi^4$ interacting model}
	\label{figTSE1}
\end{figure}

The self-energy is calculated by using the Feynman rules in 
TFD \cite{henning1,fynrule}. In the thermal doublet notation 
the Feynman rule for the four-point scalar vertex is given by 
a vector like coupling constant which consists of the coupling 
constants for the non-tilde and the tilde fields. 
Thus the following factor is assigned to the $\phi^4$ vertex, 
\begin{eqnarray}
\lambda^{\alpha} = \lambda
\left(
   \begin{array}{c}
	1 \\
	-1 \\
   \end{array}
\right ).	\label{TCC0}
\end{eqnarray}

In Fig.~\ref{figTSE1} we show the lowest order Feynman diagram 
which contributes to the time evolution equation for the neutral scalar field. 
According to the thermal propagator (\ref{ThermPro0}) 
and the coupling constant (\ref{TCC0}), the self-energy is given by
\begin{eqnarray}
&&i\Sigma_B^{\gamma_1\gamma_2}(t_{z_1}-t_{z_2};\Vec{p})  \nonumber \\
&&= -\frac{\lambda^2}{3!}\int \frac{d^2 \Vec{k}_1}{(2\pi)^2}
	\frac{d^2 \Vec{k}_2}{(2\pi)^2}
	D_0^{\gamma_1 \gamma_2}(t_{z_1}-t_{z_2},\Vec{k}_1)
	D_0^{\gamma_1 \gamma_2}(t_{z_1}-t_{z_2},\Vec{k}_2)
	D_0^{\gamma_1 \gamma_2}(t_{z_1}-t_{z_2},\Vec{q}), \nonumber \\ \label{BoseSE} 
\end{eqnarray}
with $\Vec{q} \equiv\Vec{p}-\Vec{k}_1-\Vec{k}_2$. 
Substituting the explicit expression for the thermal propagator (\ref{ThermPro0}), we obtain 
\begin{eqnarray}
&& i\Sigma_B^{\gamma_1\gamma_2}(t_{z_1}-t_{z_2};\Vec{p}) \nonumber \\
&&=-\frac{\lambda^2}{3!} \sum^{2}_{i1 =1}\sum^{2}_{i2 =1}\sum^{2}_{i3 =1}
\int \frac{d^2\Vec{k}_1}{(2\pi)^2} \frac{d^2\Vec{k}_2}{(2\pi)^2}
\frac{1}{8\omega_{k_1}\omega_{k_2}\omega_{q}} \label{BoseSelfE1} \\
\ \ \ &&\times \Biggl[ \theta( t_{z_1}-t_{z_2} ) 
e^{i(E_{k_1,i1}+E_{k_2,i2}+E_{q,i3})(t_{z_1}-t_{z_2})} \nonumber \\
&& \times \left(
   \begin{array}{cc}
	f_{k_1,i_1,a}(t_{z_2}) f_{k_2,i_2,a}(t_{z_2}) f_{q,i_3,a}(t_{z_2})
 & -f_{k_1,i_1,b}(t_{z_2}) f_{k_2,i_2,b}(t_{z_2}) f_{q,i_3,b}(t_{z_2}) \\
	f_{k_1,i_1,a}(t_{z_2}) f_{k_2,i_2,a}(t_{z_2}) f_{q,i_3,a}(t_{z_2})
 & -f_{k_1,i_1,b}(t_{z_2}) f_{k_2,i_2,b}(t_{z_2}) f_{q,i_3,b}(t_{z_2}) \\
   \end{array}
\right) \nonumber \\
&& \ \ \ + \theta( t_{z_2}-t_{z_1} ) 
e^{-i(E_{k_1,i_1}+E_{k_2,i_2}+E_{q,i_3})(t_{z_1}-t_{z_2})} \nonumber \\
&& \times \left(
   \begin{array}{cc}
	f_{k_1,i_1,a}(t_{z_1}) f_{k_2,i_2,a}(t_{z_1}) f_{q,i_3,a}(t_{z_1})
 & -f_{k_1,i_1,a}(t_{z_1}) f_{k_2,i_2,a}(t_{z_1}) f_{q,i_3,a}(t_{z_1}) \\
	f_{k_1,i_1,b}(t_{z_1}) f_{k_2,i_2,b}(t_{z_1}) f_{q,i_3,b}(t_{z_1})
 & -f_{k_1,i_1,b}(t_{z_1}) f_{k_2,i_2,b}(t_{z_1}) f_{q,i_3,b}(t_{z_1}) \\
   \end{array}
\right) 
\Biggr], \nonumber
\end{eqnarray}
where 
\begin{eqnarray}
&& E_{q,1}\equiv\omega_q,\ \ E_{q,2} \equiv -\omega_q, \nonumber \\
&& f_{q,1,a}(t) \equiv n_{q}(t),\ \ \ f_{q,2,a}(t) \equiv 1+n_{q}(t), \\
&& f_{q,1,b}(t) \equiv 1+n_{q}(t),\ \ \ f_{q,2,b}(t) \equiv n_{q}(t). \nonumber
\end{eqnarray}
We notice that unperturbed Bogoliubov parameters appear in the internal lines. 

The time dependent part in Eq.~(\ref{BoseSelfE1}) has the following form, 
\begin{eqnarray}
  V(t-t^\prime) = \theta ( \pm (t-t^\prime )) {\rm e}^{\pm iW(t-t^\prime)}.
\end{eqnarray}
It is rewritten in a Fourier integral form, 
\begin{eqnarray}
  V(t-t^\prime) 
= -i \int \frac{dp_0}{2\pi} \frac{1}{p_0 \mp W\mp i\varepsilon}
  {\rm e}^{-ip_0(t-t^\prime)}.
\label{vtfi}
\end{eqnarray}
Above $p_0$ integral is simplified in the on-shell approximation. 
We set $p_0$ on the denominator in Eq.~(\ref{vtfi}) to the on-shell value, 
$p_0=\omega_p$\cite{umezawa1}. 
Thus we can perform the Fourier integration and obtain 
\begin{eqnarray}
  V(t-t^\prime)
 = - i \delta(t-t^\prime) \frac{1}{\omega_p \mp W \mp i\varepsilon}.
\end{eqnarray}
It is decomposed into the real and the imaginary parts. 
\begin{eqnarray}
  V(t-t^\prime) 
  = -i\delta (t-t^\prime) 
  \left( \mbox{P}\frac{1}{\omega_p \mp W} \pm 2\pi i\delta^{0}(\omega_p-W) \right) ,
 \label{V-delta}
\end{eqnarray}
where P denotes the principal part of $1/(\omega_p \mp W)$. 
Under the on-shell approximation the 2-loop thermal self-energy 
(\ref{BoseSelfE1}) simplifies to 
\begin{eqnarray}
&&i\Sigma_B^{\gamma_1\gamma_2}(t-t^\prime ;\Vec{p}) \nonumber \\
&&= -i \frac{\lambda^2}{3!} \delta (t-t^\prime) 
	\sum^{2}_{i_1 =1}\sum^{2}_{i_2 =1}\sum^{2}_{i_3 =1}
 \int \frac{d^2 \Vec{k}_1}{(2\pi)^2} \frac{d^2 \Vec{k}_2}{(2\pi)^2}
 \frac{1}{8\omega_{k_1} \omega_{k_2} \omega_{q}} \nonumber \\
&&\times \Biggl[
  \frac{1}{\omega_p + E_{k_1,i_1} + E_{k_2,i_2} + E_{q,i_3} + i\epsilon}
 \nonumber \\
&& \times \left(
   \begin{array}{cc}
	f_{k_1,i_1,a}(t^\prime) f_{k_2,i_2,a}(t^\prime) f_{q,i_3,a}(t^\prime)
 & -f_{k_1,i_1,b}(t^\prime) f_{k_2,i_2,b}(t^\prime) f_{q,i_3,b}(t^\prime) \\
	f_{k_1,i_1,a}(t^\prime) f_{k_2,i_2,a}(t^\prime) f_{q,i_3,a}(t^\prime)
 & -f_{k_1,i_1,b}(t^\prime) f_{k_2,i_2,b}(t^\prime) f_{q,i_3,b}(t^\prime) \\
   \end{array}
\right) \nonumber \\
&& \ \ \ -  \frac{1}{\omega_p - E_{k_1,i_1} - E_{k_2,i_2} - E_{q,i_3} - i\epsilon}
 \nonumber \\
&& \times \left(
   \begin{array}{cc}
	f_{k_1,i_1,a}(t) f_{k_2,i_2,a}(t) f_{q,i_3,a}(t)
 & -f_{k_1,i_1,a}(t) f_{k_2,i_2,a}(t) f_{q,i_3,a}(t) \\
	f_{k_1,i_1,b}(t) f_{k_2,i_2,b}(t) f_{q,i_3,b}(t)
 & -f_{k_1,i_1,b}(t) f_{k_2,i_2,b}(t) f_{q,i_3,b}(t) \\
   \end{array}
\right) \Biggr]. \nonumber \\
	\label{BoseSelfEOnShl1}
\end{eqnarray}

The perturbed propagator, $D_H$, is necessary 
to evaluate the right-hand side 
in Eqs.~(\ref{BltzEqR-2}) and (\ref{BltzEqA-2}). 
Here we drop the higher order corrections and 
use the unperturbed propagator, $D_0$, 
instead of the perturbed one, $D_H$. 
Substituting the thermal self-energy (\ref{BoseSelfE1}) 
into the time evolution equations for the Bogoliubov parameter, 
(\ref{BltzEqR-2}) and (\ref{BltzEqA-2}) and replacing the perturbed propagator 
with the unperturbed one, 
we derive a equation with the structure of the Boltzmann equation 
for the $\lambda\phi^4$ interaction model. 
We will call it Boltzmann equation, 
\begin{eqnarray}
&&\dot{\nu}_p (t_x) =
 \frac{\lambda^2}{3!} \sum^{2}_{i1 =1}\sum^{2}_{i2 =1}\sum^{2}_{i3 =1}
 \int_{-\infty}^{t_x} dt_s
 \int \frac{d^2\Vec{k}_1}{(2{\pi})^2} \frac{d^2\Vec{k}_2}{(2{\pi})^2}
	\frac{1}{16\omega_{p} \omega_{k_1} \omega_{k_2} \omega_{q} }	\nonumber \\
&&\times {\rm cos} \bigl\{ (-\omega_{p}+E_{k_1,i_1}+E_{k_2,i_2}+E_{q,i_3}) (t_x-t_s)
  \bigr\}	\nonumber \\
&&\ \ \ \times \Bigl[ (1+n_{H,p}(t_s))
	 f_{k_1,i_1,a}(t_s) f_{k_2,i_2,a}(t_s) f_{q,i_3,a}(t_s) \nonumber \\
&&~~  - n_{H,p}(t_s) f_{k_1,i_1,b}(t_s) f_{k_2,i_2,b}(t_s)f_{q,i_3,b}(t_s) \Bigr].
	\label{BSDOffShl}
\end{eqnarray}
It should be noticed that both Eqs.~(\ref{BltzEqR-2}) and 
(\ref{BltzEqA-2}) provide the same expression (\ref{BSDOffShl}). 
In the on-shell approximation the principal part in Eq.~(\ref{V-delta}) 
is canceled out from the time evolution equation. Thus the Boltzmann 
equation (\ref{BSDOffShl}) reduces to 
\begin{eqnarray}
&&\dot{\nu}_p (t_x) =
 \frac{\lambda^2}{3!} \sum^{2}_{i_1 =1}
 \sum^{2}_{i_2 =1} \sum^{2}_{i_3 =1}
 \int \frac{d^2\Vec{k}_1}{(2{\pi})^2} \frac{d^2\Vec{k}_2}{(2{\pi})^2} 
 \frac{\pi}{8\omega_{p} \omega_{k_1} \omega_{k_2} \omega_{q} } \nonumber \\
&& \times \delta (\omega_{p} - E_{k_1,i_1} - E_{k_2,i_2} - E_{q,i_3})
  \nonumber \\
&& \times   \Bigl[(1+n_{H,p}(t_x))
	 f_{k_1,i_1,a}(t_x) f_{k_2,i_2,a}(t_x) f_{q,i_3,a}(t_x) \nonumber \\
&&~~  - n_{H,p}(t_x)f_{k_1,i_1,b}(t_x) f_{k_2,i_2,b}(t_x)f_{q,i_3,b}(t_x)  \Bigr] .
	\label{BSDOnShl}
\end{eqnarray}
The right-hand side of this equation contains the delta 
function for the energy conservation and the statistical factors. 
It corresponds to the collision term in the Boltzmann equation. 

\section{Numerical analysis of the Boltzmann equation}
The Boltzmann equation (\ref{BSDOffShl}) describes the time 
evolution of the particle number distribution. We numerically 
solve it starting from the Bose distribution with 
the temperature, $T$, and the mass, $m_0$, 
\begin{equation}
 n_p(t=0) = \frac{1}{ e^{ \sqrt[]{\Vec{p}^2+m_0^2} /T }-1 }.
\end{equation}
It is assumed that the scalar mass suddenly changes 
from $m_0$ to $m$. Then the state is no longer in 
equilibrium. Below we set parameters to 
$m_0=5\times 10^{-2} \mu$, 
$m=4\times 10^{-2} \mu$, $T=\mu$ and $\lambda=\mu$, 
with an arbitrary mass scale, $\mu$. 
It is expected that the distribution function, 
$n_p(t)$, approaches to the Bose distribution, 
$n_{f,p}$, with temperature, $T_f$. The final 
state temperature can be estimated at $T_f=0.988\mu$ 
from the energy conservation law. 

First we numerically solve the Boltzmann equation 
(\ref{BSDOnShl}) under the on-shell approximation. 
It is the differential equation in terms of the time 
variable. We employ the fourth order Runge-Kutta 
algorithm to solve it. In each steps of Runge-Kutta 
algorithm the momentum integral is performed by the 
second order Simpson integral.  The radial 
part of the momentum integral is cut off 
at the scale, $\Lambda=20\mu$.  In order to reduce the 
numerical error the integral interval is divided into 
two sub intervals, $[0,0.3\Lambda]$ and 
$[0.3\Lambda, \Lambda]$. 
The Simpson's rule is applied to each sub interval. 
The scattering with the particle in the final equilibrium state 
is introduced through the self-energy. 

For the numerical analysis we define the deviation from 
the final equilibrium distribution with the temperature, $T_f$, by 
\begin{eqnarray}
\delta n_p(t) = n_p(t) - n_{f,p}. \label{delN}
\end{eqnarray}
We expand the distribution function $f(t)$ in Eq.~(\ref{BSDOnShl}) in terms of $\delta n$, 
\begin{eqnarray}
&& f_{q,i,a(b)}(t)= f^f_{q,i,a(b)} + O(\delta n (t)), 
\end{eqnarray}
where $i=1,2$ and 
\begin{eqnarray}
&& f_{q,1,a}^f = f_{q,2,b}^f  = n_{f,q},\\
&& f_{q,2,a}^f = f_{q,1,b}^f = 1+n_{f,q}.
\end{eqnarray}

In our setup the initial state temperature, $T=\mu$, 
is close to the expected final state temperature, $T_f=0.988\mu$. 
Then we keep only the leading order $\delta n$ expansion in numerical calculations. 
The Bogoliubov parameters is fixed to $n_{f,p}$ for 
the internal lines in the self-energy. We solve the 
time evolution of the one for the external line, $n_H$. 

\begin{figure}[htb]
	\begin{center}
	\includegraphics{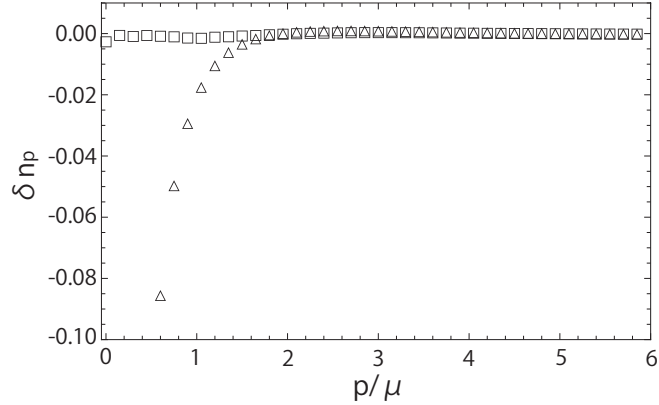}
	\end{center}
	\caption{ Behavior of $\delta n_p(t)$ in the on-shell 
        approximation (\ref{BSDOnShl}). The triangle and square
        points show $\delta n_p(t=0)$ and 
        $\delta n_p(t=1.0\times10^{2}\mu^{-1})$, respectively. }
	\label{FigdBnOnShl}
\end{figure}

\begin{figure}[htb]
	\begin{center}
	\includegraphics{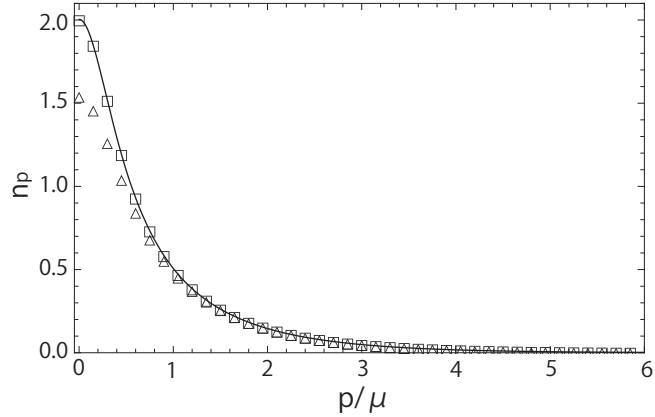}
	\end{center}
	\caption{ Behavior of $n_p(t)$ in the on-shell 
        approximation (\ref{BSDOnShl}). The triangle and square
        points show $n_p(t=0)$ and 
        $n_p(t=1.0\times10^{2}\mu^{-1})$, respectively. The solid
        line represents the Bose distribution at $T=T_f$. }
	\label{FigBnOnShl}
\end{figure}

Evaluating the Boltzmann equation (\ref{BSDOnShl}) in above 
assumptions, we obtain the time evolution of the distribution function. 
In Fig.~\ref{FigdBnOnShl} the behavior of $\delta n_p(t)$ is 
shown as a function of the momentum $p$. At $t=0$ the particle 
number density with a lower momentum is much smaller than the 
estimated final one. We also draw the behavior of the particle 
number distribution, $n_p(t)$, in Fig.~\ref{FigBnOnShl}. 
As is expected, the particle number distribution approaches the 
equilibrium distribution with the temperature, $T_f$. 

To evaluate the off-shell mode contribution we rewrite Eq.~(\ref{BSDOffShl}) as 
\begin{eqnarray}
&&\int_{-\infty}^{t_x} dt_s \ \dot{\nu}(t_s) =
 \frac{\lambda^2}{3!} \sum^{2}_{i1 =1}\sum^{2}_{i2 =1}\sum^{2}_{i3 =1}
 \int_{-\infty}^{t_x} dt_s
 \int \frac{d^2\Vec{k}_1}{(2{\pi})^2} \frac{d^2\Vec{k}_2}{(2{\pi})^2}
	\frac{1}{16\omega_{p} \omega_{k_1} \omega_{k_2} \omega_{q} }	\nonumber \\
&&\times \frac{{\rm sin} \bigl\{ (-\omega_{p}+E_{k_1,i1}+E_{k_2,i2}+E_{q,i3}) (t_x-t_s)
  \bigr\}}
 { -\omega_{p}+E_{k_1,i1}+E_{k_2,i2}+E_{q,i3} }	\nonumber \\
&&\times \Bigl[ (1+n_{H,p}(t_s)) f_{k_1,i1,a}(t_s) f_{k_2,i2,a}(t_s) f_{q,i3,a}(t_s)
 \nonumber \\
 &&~~ - n_{H,p}(t_s) f_{k_1,i1,b}(t_s) f_{k_2,i2,b}(t_s)f_{q,i3,b}(t_s) \Bigr].
	\label{BSDOffShlKai}
\end{eqnarray}
The Boltzmann equation (\ref{BSDOffShl}) is reproduced by 
differentiating both the sides of this equation in terms of $t_x$. 
This equation is satisfied for an arbitrary $t_x$. 
Thus we obtain the differential equation, 
\begin{eqnarray}
&&\ \dot{\nu}(t_s) =
 \frac{\lambda^2}{3!} \sum^{2}_{i_1 =1}\sum^{2}_{i_2 =1}\sum^{2}_{i_3 =1}
 \int \frac{d^2\Vec{k}_1}{(2{\pi})^2} \frac{d^2\Vec{k}_2}{(2{\pi})^2}
	\frac{1}{16\omega_{p} \omega_{k_1} \omega_{k_2} \omega_{q} }	\nonumber \\
&&\times \frac{{\rm sin} \bigl\{ (-\omega_{p}+E_{k_1,i_1}+E_{k_2,i_2}+E_{q,i_3}) (t_x-t_s)
  \bigr\}}
 { -\omega_{p}+E_{k_1,i_1}+E_{k_2,i_2}+E_{q,i_3} }	\nonumber \\
&&\times \Bigl[ (1 + n_{H,p}(t_s))
 f_{k_1,i_1,a}(t_s) f_{k_2,i_2,a}(t_s) f_{q,i_3,a}(t_s) \nonumber \\
&&~~  - n_{H,p}(t_s) f_{k_1,i1,b}(t_s) f_{k_2,i2,b}(t_s)f_{q,i3,b}(t_s) \Bigr].
	\label{BSDOffShlKai2}
\end{eqnarray}
The time integral is dropped. Therefore the same numerical 
algorithm for the on-shell approximation can be applied to 
Eq.~(\ref{BSDOffShlKai2}). The time variable, 
$t_x$, in Eq.~(\ref{BSDOffShlKai2}) represents the time 
interval from the initial moment. Here we set 
$t_x=1.0\times 10^3\mu^{-1}$. At the limit, 
$t_x\rightarrow \infty$, the off-shell mode contribution disappears. 

\begin{figure}[htb]
	\begin{center}
	\includegraphics{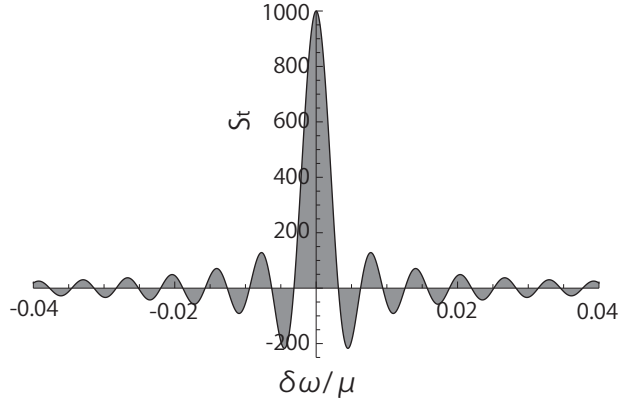}
	\end{center}
	\caption{Behavior of $S_t(\delta\omega)$ 
                for $t_x-t_s=1000\mu^{-1}$.}
	\label{FigSanitsu}
\end{figure}

The integral kernel in Eq.~(\ref{BSDOffShlKai2}) is proportional to 
\begin{eqnarray}
S_{t}(\delta \omega) \equiv \frac{{\rm sin}\{ \delta \omega (t_x-t_s)\} }{\delta \omega}.
\end{eqnarray}
As is shown in Fig.~\ref{FigSanitsu}, it has a peak at the on-shell limit, 
$\delta \omega\rightarrow 0$, and frequently oscillates for a larger $\delta \omega$. 
Thus it is enough to evaluate the integral near the on-shell limit. 
We restrict the integral interval, $-12\pi\le \delta\omega(t_x-t_s)\le 12\pi$. 

\begin{figure}[htb]
	\begin{center}
	\includegraphics{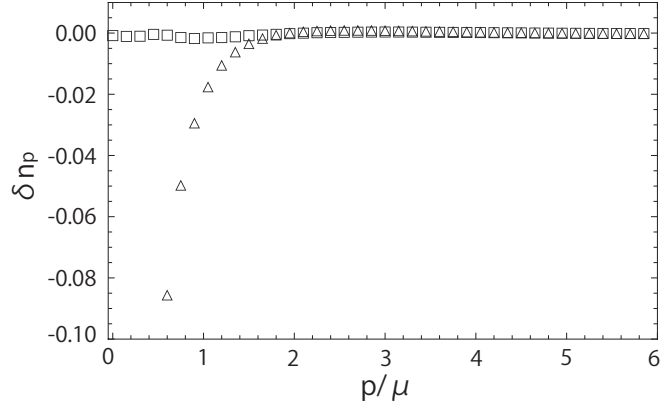}
	\end{center}
	\caption{ Behavior of $\delta n_p(t)$ with the off-shell
        contribution (\ref{BSDOffShlKai2}).
        The triangle and square
        points show $\delta n_p(t=0)$ and 
        $\delta n_p(t=1.0\times10^{2}\mu^{-1})$, respectively. }
	\label{FigdBnOffShl}
\end{figure}

\begin{figure}[htb]
	\begin{center}
	\includegraphics{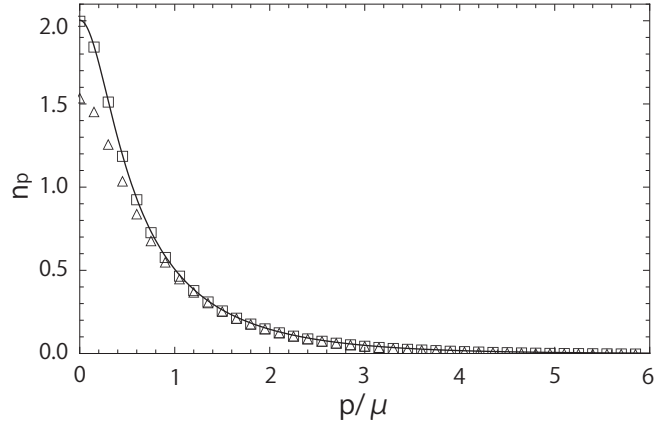}
	\end{center}
	\caption{ Behavior of $n_p(t)$ with the off-shell
        contribution (\ref{BSDOffShlKai2}).
        The triangle and square
        points show $n_p(t=0)$ and 
        $n_p(t=1.0\times10^{2}\mu^{-1})$, respectively. The solid
        line represents the Bose distribution at $T=T_f$. }
	\label{FigBnOffShl}
\end{figure}

We numerically evaluate the Boltzmann equation (\ref{BSDOffShlKai2}) 
in above approximation and obtain the time evolution of the 
distribution function with the off-shell mode contribution. 
The behavior of $\delta n_p(t)$ and $n_p(t)$ is illustrated 
as a function of the momentum $p$ in Figs.~\ref{FigdBnOffShl} 
and \ref{FigBnOffShl}, respectively. 
Hence, a similar behavior is observed for Eqs.~(\ref{BSDOnShl}) 
and (\ref{BSDOffShlKai2}). It seems to be difficult to distinguish 
the contribution from the off-shell mode in these figures. 

To evaluate the contribution from the off-shell mode we calculate 
the relaxation time, $\tau_p$. According to the linear response theory, 
we define it by 
\begin{eqnarray}
\tau_p \equiv -\frac{ \delta n_p} {\delta \dot{n_p}}.
\label{rtime}
\end{eqnarray}
It can be calculated by solving the Boltzmann equations 
(\ref{BSDOnShl}) and (\ref{BSDOffShlKai2}). 

\begin{figure}[htb]
	\begin{center}
	\includegraphics{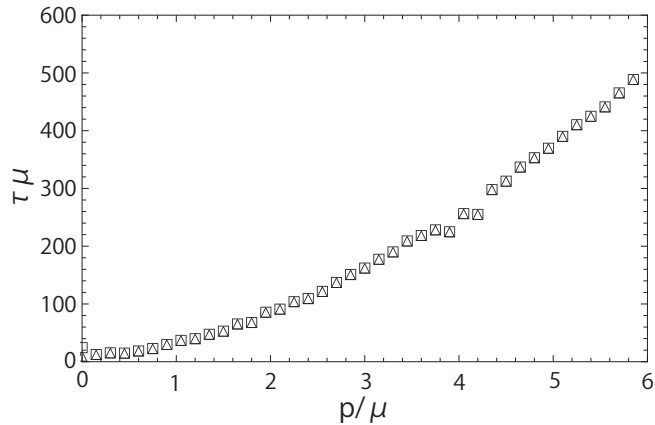}
	\end{center}
	\caption{ The relaxation time, $\tau_p$, in the on-shell 
        approximation. The triangle and square
        points show $\tau(t=0)$ and 
        $\tau(t=1.0\times10^{2}\mu^{-1})$, respectively. }
	\label{FigTauOnShl}
\end{figure}

\begin{figure}[htb]
	\begin{center}
	\includegraphics{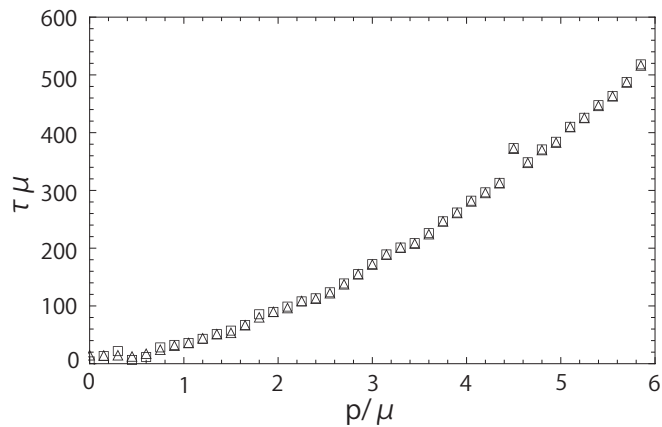}
	\end{center}
	\caption{ The relaxation time, $\tau_p$, with the off-shell 
        contribution. The triangle and square
        points show $\tau(t=0)$ and 
        $\tau(t=1.0\times10^{2}\mu^{-1})$, respectively. }
	\label{FigTauOffShl}
\end{figure}

We plot the behavior of the relaxation time, $\tau_p$, as a function of 
the momentum, $p$, in Fig.~\ref{FigTauOnShl} under the on-shell 
approximation. The off-shell contribution is included in 
Fig.~\ref{FigTauOffShl}. The relaxation time 
is almost constant with respect to the time variable, $t$. 
A longer relaxation time is observed for a higher momentum mode. 
As is shown in Figs.~\ref{FigBnOnShl} and \ref{FigBnOffShl} 
the distribution function $n(t)$ is close to the final 
equilibrium state at $t=1.0\times10^{2}\mu^{-1}$. 
It is consistent with the behavior of the relaxation time. 
For $p\lesssim 2.0\mu$ the relaxation time is smaller than $1.0\times10^{2}\mu^{-1}$. 
The particle number distribution for a higher momentum mode is 
close to the final equilibrium state at $t=0$ and slowly approaches the final state. 

A small but stable discrepancy is observed for some points 
around $p\sim 4\mu$ in Figs.~\ref{FigTauOnShl} and \ref{FigTauOffShl}. 
It corresponds to the numerical error which is produced 
due to a finite step size in the Simpson integral. 

\begin{figure}[htb]
\begin{center}
\includegraphics{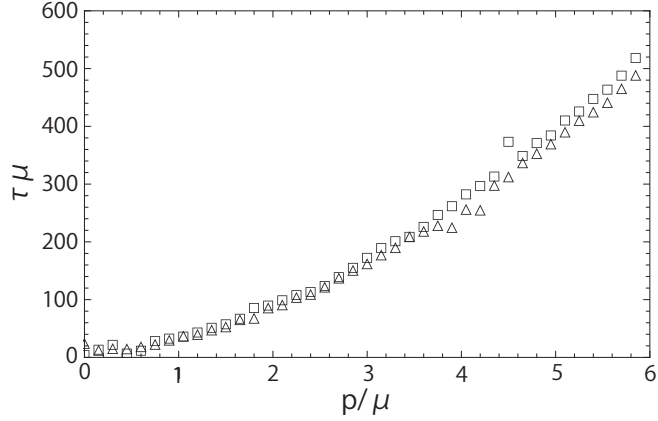}
\end{center}
\caption{ Behavior of the relaxation time, $\tau_p$.
        The triangle and square
        points show the relaxation time without and with
        the off-shell contribution, respectively. }
\label{FigTauOnShlOffShl}
\end{figure}

In Fig.~\ref{FigTauOnShlOffShl} both of the results are drawn in the same figure. 
The off-shell mode contribution does not modify the relaxation time for a lower momentum. 
For $p>3\mu$ we observe a shorter relaxation time in 
the on-shell approximation. Since the off-shell contribution 
destabilizes the particle distribution near the equilibrium state, 
it slightly increases the relaxation time. A different contribution 
from the off-shell mode is expected for a system far from the 
equilibrium state. Because of a non-negligible numerical error it is 
difficult to apply our analysis to such a system. Some improvements 
of the numerical algorithm are necessary to extend our analysis for a general case. 

\section{Conclusion}
We have investigated the time evolution of the distribution 
function for a relativistic neutral scalar field with a 
$\lambda\phi^4$ interaction. 
The NETFD is applied to the SD equation for the scalar field propagator. 
Calculating the 2-loop thermal self-energy and inserting it 
into the SD equation, we have derived the time 
evolution equation for the thermal Bogoliubov parameter. 
The equation has the same structure with the Boltzmann 
equation. Therefore the Boltzmann equation is obtained 
from the SD equation at the equal time limit. 

The Boltzmann equation consists of the collision terms 
which depend on the thermal Bogoliubov parameter and an 
oscillating coefficient with respect to the time variable. 
The oscillating coefficient has a peak for the on-shell 
case. It reduces to the delta function which shows the energy 
conservation between the collision particles in the 
on-shell approximation. 
Solving the obtained Boltzmann equation, we have evaluated 
the time evolution of the thermal Bogoliubov parameter. 
We suppose that the scalar field loses 20\% of mass 
suddenly at $t=0$. 
It is observed that the thermal Bogoliubov parameter approaches an equilibrium state. 
The relaxation time monotonically increases as a function 
of the momentum. In our setup the off-shell mode has only 
a small contribution to the particle number density. 
The off-shell mode tends to suppress the relaxation near 
the equilibrium state and slightly increases the 
relaxation time for a higher momentum region. 

We have derived the Boltzmann equation from the SD equation. 
For a non-relativistic scalar field 
the Boltzmann equation has been derived as the self-consistent 
renormalization condition based on the canonical 
quantization by H.~Umezawa and Y.~Yamanaka 
\cite{umezawa1,CQNETFD1,CQNETFD2,CQNETFD3}. 
In this case it has been shown that the thermal Bogoliubov 
parameter which satisfies the self-consistent renormalization 
condition coincides with the observed particle number 
density \cite{umezawa1,netfd1,netfd2}. 
We would like to show the correspondence between the thermal 
Bogoliubov parameter and the particle number density in 
our approach. For this purpose it is necessary to construct 
our procedure based on the canonical quantization. 

In the derivation of the time evolution equation the second 
derivative of the  Bogoliubov parameter is canceled out at 
the equal time limit. Thus we obtain the Markovian equation. 
In the calculation of the self-energy 
part we impose the internal scalar propagator to be in the 
final equilibrium state. We should improve these assumptions to generalize 
the procedure. It is also interesting to apply our analysis 
to the relativistic Dirac field. These will be the subject of 
a forthcoming paper. 

\section*{Acknowledgements}
The authors would like to thank Y.~Yamanaka, H.~Matsumoto and 
Y.~Nakamura for valuable discussions. 
Discussions during the YITP workshop 
on "Thermal Quantum Field Theories and Their 
Applications 2010" were useful to complete this work. 

\appendix
\section{diagonal elements of the SD equation}

The relationship between the perturbed and the unperturbed propagator 
is given by the SD equations (\ref{BoseSDeq0}) and (\ref{BoseSDeq0-2}). 
Here we evaluate the diagonal elements of these equations. 
Applying the Bogoliubov matrix, $B^{-1}(n_H)$ and $B(n_H)$ to 
Eq.~(\ref{BoseSDeq0}) on the left and the right sides, respectively, 
we obtain conditions for each diagonal element of the perturbed propagator. 
\begin{eqnarray}
&&D_{H,R}^{11}(t_x-t_y) + D_{H,A}^{22}(t_x-t_y) 
 =D_{0,R}^{11}(t_x-t_y)  + D_{0,A}^{22}(t_x-t_y) \nonumber \\
&&\ +\int dt_{z_1} dt_{z_2} \Bigl(D_{0,R}^{11}(t_x-t_{z_1})+D_{0,A}^{22}(t_x-t_{z_1})\Bigr)
\nonumber \\
&&~~\times i \Sigma_{R}(t_{z_1}-t_{z_2})
 \Bigl(D_{H,R}^{11}(t_{z_2}-t_y)+D_{H,A}^{22}(t_{z_2}-t_y)\Bigr), \label{FullD1} \\
&&D_{H,R}^{22}(t_x-t_y) + D_{H,A}^{11}(t_x-t_y) =
 D_{0,R}^{22}(t_x-t_y) + D_{H,A}^{11}(t_x-t_y)  \nonumber \\
&& +\int dt_{z_1} dt_{z_2} \Bigl(D_{0,R}^{22}(t_x-t_{z_1})+D_{0,A}^{11}(t_x-t_{z_1})\Bigr)
\nonumber \\
&&~~\times i \Sigma_{A}(t_{z_1}-t_{z_2})
 \Bigl( D_{H,R}^{22}(t_{z_2}-t_y)+D_{H,A}^{11}(t_{z_2}-t_y)\Bigr). \label{FullD2}
\end{eqnarray}
We suppose that the retarded and the advanced parts 
of these equations separate and impose conditions, 
\begin{eqnarray}
&&D_{H,R}^{11}(t_x-t_y)=D_{0,R}^{11}(t_x-t_y)  \label{propFull1}\\
&&\ +\int dt_{z_1} dt_{z_2} \Bigl(D_{0,R}^{11}(t_x-t_{z_1})+D_{0,A}^{22}(t_x-t_{z_1})\Bigr)
i \Sigma_{R}(t_{z_1}-t_{z_2})D_{H,R}^{11}(t_{z_2}-t_y), \nonumber \\
&&D_{H,A}^{22}(t_x-t_y)=D_{0,A}^{22}(t_x-t_y)  \label{propFull2}\\
&&\ +\int dt_{z_1} dt_{z_2} \Bigl(D_{0,R}^{11}(t_x-t_{z_1})+D_{0,A}^{22}(t_x-t_{z_1})\Bigr)
i \Sigma_{R}(t_{z_1}-t_{z_2})D_{H,A}^{22}(t_{z_2}-t_y), \nonumber  \\
&&D_{H,R}^{22}(t_x-t_y)=D_{0,R}^{22}(t_x-t_y)  \label{propFull3} \\
&&\ +\int dt_{z_1} dt_{z_2} \Bigl(D_{0,R}^{22}(t_x-t_{z_1})+D_{0,A}^{11}(t_x-t_{z_1})\Bigr)
i \Sigma_{A}(t_{z_1}-t_{z_2})D_{H,R}^{22}(t_{z_2}-t_y), \nonumber \\
&&D_{H,A}^{11}(t_x-t_y)=D_{0,A}^{11}(t_x-t_y)  \label{propFull4} \\
&&\ +\int dt_{z_1} dt_{z_2} \Bigl(D_{0,R}^{22}(t_x-t_{z_1})+D_{0,A}^{11}(t_x-t_{z_1})\Bigr)
i \Sigma_{A}(t_{z_1}-t_{z_2})D_{H,A}^{11}(t_{z_2}-t_y). \nonumber  
\end{eqnarray}
From the SD equation (\ref{BoseSDeq0-2}) we find other 
conditions for the perturbed propagator. 
\begin{eqnarray}
&&D_{H,R}^{11}(t_x-t_y) + D_{H,A}^{22}(t_x-t_y) 
 =D_{0,R}^{11}(t_x-t_y)  + D_{0,A}^{22}(t_x-t_y) \nonumber \\
&&\ +\int dt_{z_1} dt_{z_2} \Bigl(D_{H,R}^{11}(t_x-t_{z_1})+D_{H,A}^{22}(t_x-t_{z_1})\Bigr)
\nonumber \\
&&~~\times i \Sigma_{R}(t_{z_1}-t_{z_2})
 \Bigl(D_{0,R}^{11}(t_{z_2}-t_y)+D_{0,A}^{22}(t_{z_2}-t_y)\Bigr), \label{fullD3} \\
&&D_{H,R}^{22}(t_x-t_y) + D_{H,A}^{11}(t_x-t_y) =
 D_{0,R}^{22}(t_x-t_y) + D_{H,A}^{11}(t_x-t_y)  \nonumber \\
&& +\int dt_{z_1} dt_{z_2} \Bigl( D_{H,R}^{22}(t_x-t_{z_1})+D_{H,A}^{11}(t_x-t_{z_1})\Bigr)
\nonumber \\
&&~~\times i \Sigma_{A}(t_{z_1}-t_{z_2})
 \Bigl( D_{0,R}^{22}(t_{z_2}-t_y)+D_{0,A}^{11}(t_{z_2}-t_y)\Bigr). \label{fullD4}
\end{eqnarray}
In a similar manner with Eqs.~(\ref{propFull1})-(\ref{propFull4}) we divide 
Eqs.~(\ref{fullD3}) and (\ref{fullD4}), 
\begin{eqnarray}
&&D_{H,R}^{11}(t_x-t_y)=D_{0,R}^{11}(t_x-t_y)  \label{propFull5} \\
&&\ +\int dt_{z_1} dt_{z_2} D_{H,R}^{11}(t_x-t_{z_1}) i \Sigma_{R}(t_{z_1}-t_{z_2})
\Bigl( D_{0,R}^{11}(t_{z_2}-t_y)+D_{0,A}^{22}(t_{z_2}-t_y)\Bigr), \nonumber \\
&&D_{H,A}^{22}(t_x-t_y)=D_{0,A}^{22}(t_x-t_y)  \label{propFull6} \\
&&\ +\int dt_{z_1} dt_{z_2} D_{H,A}^{22}(t_x-t_{z_1}) i \Sigma_{R}(t_{z_1}-t_{z_2})
\Bigl( D_{0,R}^{11}(t_{z_2}-t_y)+D_{0,A}^{22}(t_{z_2}-t_y)\Bigr), \nonumber  \\
&&D_{H,R}^{22}(t_x-t_y)=D_{0,R}^{22}(t_x-t_y)  \label{propFull7}\\
&&\ +\int dt_{z_1} dt_{z_2} D_{H,R}^{22}(t_x-t_{z_1}) i \Sigma_{A}(t_{z_1}-t_{z_2})
\Bigl( D_{0,R}^{22}(t_{z_2}-t_y)+D_{0,A}^{11}(t_{z_2}-t_y)\Bigr), \nonumber \\
&&D_{H,A}^{11}(t_x-t_y)=D_{0,A}^{11}(t_x-t_y)  \label{propFull8} \\
&&\ +\int dt_{z_1} dt_{z_2} D_{H,A}^{11}(t_x-t_{z_1}) i \Sigma_{A}(t_{z_1}-t_{z_2})
\Bigl( D_{0,R}^{22}(t_{z_2}-t_y)+D_{0,A}^{11}(t_{z_2}-t_y)\Bigr). \nonumber  
\end{eqnarray}
The self-energy, $\Sigma_{R,A}$ has retarded and advanced 
time dependence, $\Sigma_R(t-t^\prime)=0$ for $t < t^\prime$ 
and $\Sigma_A(t-t^\prime)=0$ for $t > t^\prime$. 
The time dependence for $D_{R}$ and $D_{A}$ is given in 
Eqs.~(\ref{BoseThermPro1-1})-(\ref{BoseThermPro1}). 
Thus the quantum corrections for the perturbed propagators (\ref{propFull1})
-(\ref{propFull4}) and (\ref{propFull5})-(\ref{propFull8}) disappear 
at the equal time limit. 
As a result the perturbed propagators at the equal time limit 
are derived from the unperturbed propagators 
(\ref{BoseThermPro1-1})-(\ref{BoseThermPro1}). 
\begin{eqnarray}
&&\lim_{t_x\rightarrow t_y} D_{H,R}^{11}(t_x-t_y) 
= \lim_{t_x\rightarrow t_y} D_{H,A}^{11}(t_x-t_y) 
= \frac{1}{4\omega_p},  \label{DHeqltim1} \\
&&\lim_{t_x\rightarrow t_y} D_{H,R}^{22}(t_x-t_y) 
=\lim_{t_x\rightarrow t_y} D_{H,A}^{11}(t_x-t_y)
= -\frac{1}{4\omega_p}. \label{DHeqltim2}
\end{eqnarray}

Substituting the expressions (\ref{BoseThermPro1-1})-(\ref{BoseThermPro1}) 
for the unperturbed propagators, 
we obtain the first derivative of the perturbed propagators 
at the equal time limit, 
\begin{eqnarray}
&&\lim_{t_x\rightarrow t_y} \partial_{t_x}D_{H,R}^{22}(t_x-t_y) 
=\lim_{t_x\rightarrow t_y} \partial_{t_y}D_{H,A}^{22}(t_x-t_y) 
=\frac{i}{2} + \lim_{t_x\rightarrow t_y} \delta(t_x-t_y), 
\label{DHeqltim3}  \\
&&\lim_{t_x\rightarrow t_y} \partial_{t_x}D_{H,A}^{11}(t_x-t_y) 
=\lim_{t_x\rightarrow t_y} \partial_{t_y}D_{H,R}^{11}(t_x-t_y) 
= \frac{i}{2} - \lim_{t_x \rightarrow t_y}\delta(t_x - t_y).  
\label{DHeqltim4}
\end{eqnarray}

\end{document}